\documentclass[format=acmsmall,manuscript]{acmart}

\usepackage{subfigure}
\usepackage{hyperref}
\usepackage{calc}

\usepackage{listings}
\newcommand{\lstprotectioncode}[1]{\color{red}#1}

\let\origthelstnumber\thelstnumber

\makeatletter
\newlength{\linenumwidth} 
\newlength{\numwidth}%
\def\lst@PlaceNumber{%
  \makebox[\numwidth+1em][l]{%
    \makebox[\numwidth][r]{\normalfont\lst@numberstyle{\thelstnumber}}%
  }%
}

\newcommand*\Suppressnumber{%
  \lst@AddToHook{OnNewLine}{%
    \let\thelstnumber\relax%
     \advance\c@lstnumber-\@ne\relax%
    }%
}

\newcommand*\Reactivatenumber[1]{%
  \setcounter{lstnumber}{\numexpr#1-1\relax}
  \lst@AddToHook{OnNewLine}{%
   \let\thelstnumber\origthelstnumber%
   \refstepcounter{lstnumber}
  }%
}
\makeatother

\acmJournal{FACMP}
\acmVolume{x}
\acmNumber{x}
\acmArticle{x}
\acmMonth{0}

\setcopyright{acmcopyright}
\copyrightyear{202x}
\acmYear{202x}
\acmDOI{10.1145/xxxxxxxx.xxxxxxx}

\begin{document}

\title{Flexible Software Protection}

\author{Jens Van den Broeck}
\email{jens.vandenbroeck@ugent.be}
\orcid{0002-8373-2949}
\author{Bart Coppens}
\email{bart.coppens@ugent.be}
\orcid{0002-7628-9264}
\author{Bjorn De Sutter}
\email{bjorn.desutter@ugent.be}
\orcid{0003-0317-2089}
\affiliation{%
  \institution{Department of Electronics and Information Systems, Ghent University}
  \streetaddress{Technologiepark-Zwijnaarde 126}
  \city{Ghent}
  \country{Belgium}
  \postcode{9052}
}

\renewcommand{\shortauthors}{Van den Broeck et al.}

\begin{abstract}
To counter software reverse engineering or tampering, software obfuscation tools can be used. 
However, such tools to a large degree hard-code how the obfuscations are deployed. They hence lack resilience and stealth in the face of many attacks.
To counter this problem, we propose the novel concept of flexible obfuscators, which implement protections in terms of data structures and APIs already present in the application to be protected. The protections are hence tailored to the application in which they are deployed, making them less learnable and less distinguishable.
In our research, we concretized the flexible protection concept for opaque predicates. We designed an interface to enable the reuse of existing data structures and APIs in injected opaque predicates, we analyzed their resilience and stealth, we implemented a proof-of-concept flexible obfuscator, and we evaluated it on a number of real-world use cases.
This paper presents an in-depth motivation for our work, the design of the interface, an in-depth security analysis, and a feasibility report based on our experimental evaluation.
The findings are that flexible opaque predicates indeed provide strong resilience and improved stealth, but also that their deployment is costly, and that they should hence be used sparsely to protect only the most security-sensitive code fragments that do not dominate performance.  
Flexible obfuscation therefor delivers an expensive but also more durable new weapon in the ever ongoing software protection arms race.  

\end{abstract}



\maketitle

\section{Introduction}
\label{sec:intro}
Many commercial software applications are susceptible to man-at-the-end (MATE) attacks, such as reverse engineering or tampering. Software developers want to make sure that confidential data and code in their application cannot easily be extracted and that their product is used the way it is meant to be, i.e., that the integrity of the code and of the data on which the code operates is guaranteed. Hence, they want to mitigate as many relevant attacks as possible. To hamper such attacks, software protection techniques such as code obfuscation and remote attestation are deployed~\cite{collbergbook}.

The research presented here focuses mostly on control flow obfuscation techniques, which extend and replace the original control flow of the program by artificially more complex forms that cannot easily be reversed. As a result, the time required to analyse the application increases, and the precision and recall of the attackers' analyses can be decreased. Examples of control flow obfuscations are control flow flattening~\cite{wang2000software}, branch functions~\cite{linn2003obfuscation}, and bogus control flow insertion by means of opaque predicates (OPs)~\cite{myles2006software,majumdar2006manufacturing}.
 
Adversaries will often try to analyse the applied protections to revert them or work around them, for which they can deploy several strategies. A first attack strategy to automate the deobfuscation is to identify fragments that are known a priori to be overly artificially complex replacements of simpler control flow constructs, and then replacing them by the known simpler variations. This strategy can build on techniques such as pattern matching~\cite{ngo2007detecting}, symbolic execution~\cite{yadegari2015symbolic}, and abstract interpretation~\cite{dalla2006opaque}. For some techniques, off-the-shelf extensible frameworks are available that attackers can easily customize~\cite{idapro,grap}. A key feature is that the attackers know in advance what fragments they are looking for and at what abstraction level.

A second strategy builds on the assumption that the code implementing protections is often superfluous and unnecessarily complex with respect to the semantics of the protected software, and that such protection code stands out from original program code in detectable ways that allow at once the deobfuscation and simplification of obfuscated control flow. This is the overall strategy of generic deobfuscation, simplifying code that it identifies as behaving quasi-invariantly in execution traces (i.e., producing the same result every time they are executed) and then removes code that it identifies as not contributing to the observed semantics of the traces, i.e., the observed input-output relation~\cite{yadegari15generic}. 

Fundamentally, the first strategy builds on a priori protection-specific knowledge, while the second strategy builds on the observation that protection code functions differently from original application code. In empirical research, we also observed that when attackers face programs that are small enough for their manual analysis scalability, some of them skip automated tools to detect certain fragments and instead manually scan and analyse the code in search for the relevant protection fragments~\cite{emse2019}.

Protection code is injected by software protection tools, of which a variety exists~\cite{tigress,diablo2005,llvmobf}.
While the tools offer a convenient way to deploy software protections, most of them are limited in functionality by design. Concretely, they can only deploy a pre-determined set of protections and the injected protection code only spans a limited number of code patterns, which differ from the patterns in the original software and which most often confirm the observations exploited by Yadegari et al.\ in their generic deobfuscation technique. In other words, the injected code is learnable such that it becomes attackable with the first discussed strategy, and it differs fundamentally from the original code such that it becomes attackable with the second strategy as well. 

To block both attack strategies, we propose the radical shift of no longer supporting and implementing protections in a pre-determined manner hard-coded into protection tools, but instead to make those protection tools implement the injected protections by means of code fragments already present and used in the original software to be protected. In other words, we propose to reuse already present application code for implementing the protections. 

When this paradigm shift is done well, we project several potential benefits. A first consequence is that the code that implements the protections can vary from one protected application to another. Attackers can hence not learn protections as easily a priori, which will reduce the effectiveness of the first automated attack strategy. The second consequence is that the code fragments used for implementing the protections will by definition contribute to the original program semantics, as they are still executed as part of its original computations. The third consequence is that those fragments will not display quasi-invariant behavior, as they are executed in multiple contexts that provide them different inputs for which they produce different outputs. The latter two consequences will help to mitigate the second automated attack strategy of generic deobfuscation and other attacks. The resilience of protections will hence be improved by making them more stealthy for the discussed attack strategies and tools.  

A fourth consequence will be that the code of the protected application can no longer be cleanly partitioned into application code and protection code, as the reused code fragments play a role in both parts. We know from software engineering paradigms such as separation of concerns that code is easier to comprehend and to maintain when the software components have clearly defined roles and interfaces in a program. When some code fragments start playing multiple roles, as we propose, human analysis and comprehension of the code might suffer. The potency of the protections, in particular obfuscations, might therefore potentially be improved as well. Needless to say, when code fragments implement both original application functionality and protection functionality, tampering with those fragments to bypass or undo protections also becomes harder as the tampering then inherently also impacts the original program semantics. 

In our research, we studied the proposed shift by instantiating it for OPs. We studied options to encode OPs, we studied the feasibility of reusing code in existing real-world applications, and we implemented a prototype implementation to validate our work. This paper reports on our research, and offers the following contributions:
\begin{enumerate}
   \item We pitch the idea of reusing application code for flexibly implementing software protections.
   \item As a case study, we discuss how this idea can be made concrete for flexible OPs.
   \item We present a meta-API to enable automated reuse of application code in flexible OPs.
   \item As a proof-of-concept, we present an open-source, automated tool to inject flexible OPs.
   \item We perform an in-depth security analysis, focusing on the resilience of flexible OPs.
   \item We present an experimental validation and evaluation on a number of real-world programs.
\end{enumerate}

The paper is structured along the lines of these contributions.

\section{Code reuse for software protections}
\label{sec:reuse}
To hamper automated deobfuscation, we propose to reuse existing application code for implementing software protections. The injected protection code will then look and behave more similar to the original application code.
This will make it harder for attackers to identify the protection code and to analyze and alter it with targeted techniques.

\begin{figure}[t]
  \centering
  \includegraphics[width=0.63\linewidth]{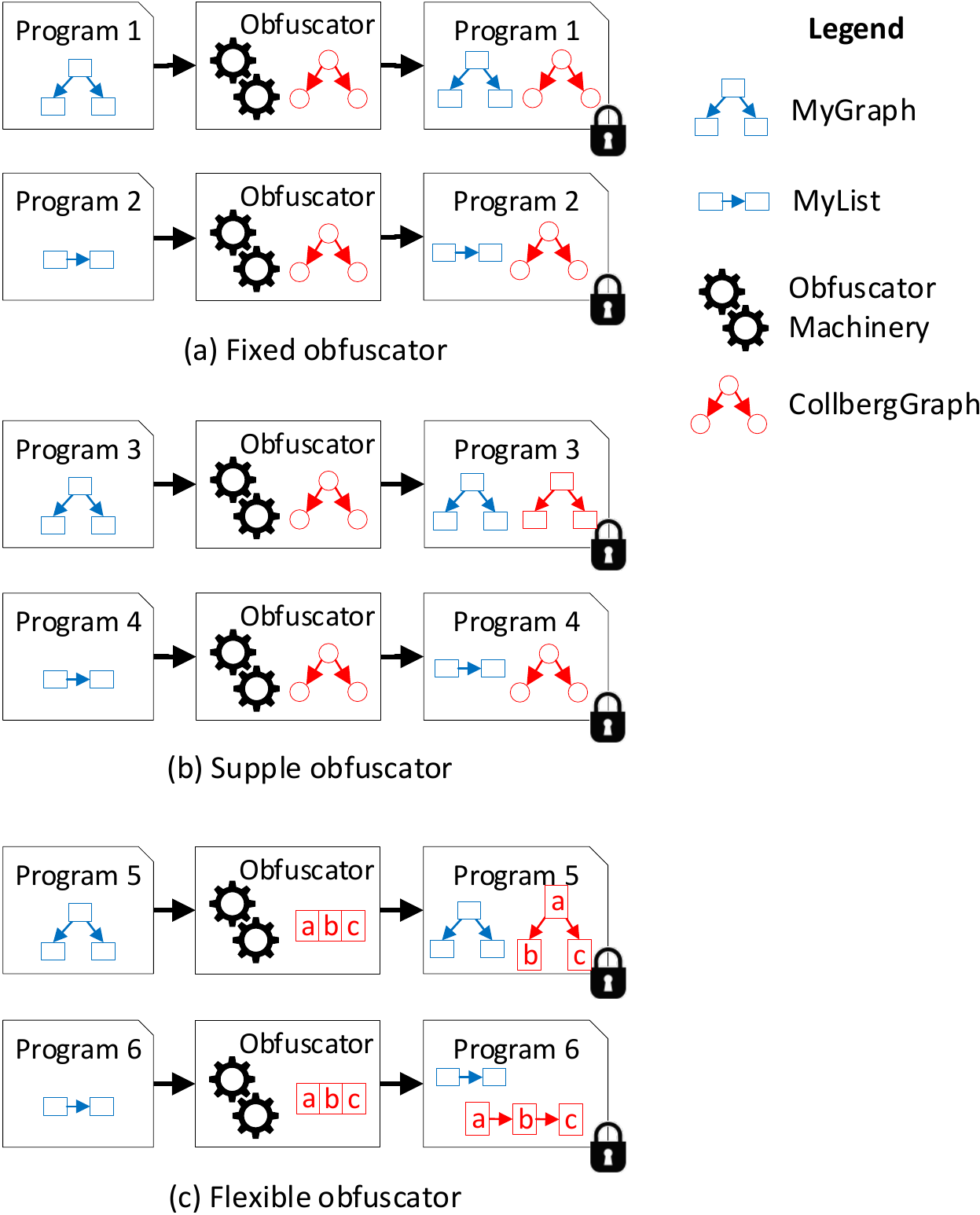}
  \caption{Overview of obfuscator designs.}
  \label{fig:obfuscator}
  \Description{Fully described in the text.}
\end{figure}

Conceptually, we propose the radical shift of no longer constructing protection tools with a predetermined and hard-coded set of data structures to implement protections, but instead making protection tools reuse existing application code to instantiate and manipulate data structures present in the unprotected application. Figure~\ref{fig:obfuscator}(a) illustrates the deployment of a traditional, fixed obfuscator that protects two programs that use the different data structures shown in blue.
In this example, the obfuscator injects code to deploy Collberg's alias-based OPs~\cite{collberg98}, which are based on comparisons of pointers to elements in graphs. To inject that code, the obfuscator incorporates a code transformation engine, as well as a fixed implementation of a graph data structure, e.g., in the form of a CollbergGraph class, which is shown in red.

In general, and while possibly being the subject of some syntactic diversification, the code injected by a traditional obfuscator consists of fixed implementations of protection techniques that deploy fixed data structures that were determined by the obfuscator developer at the design time of the obfuscator.
Like programs 1 and 2, all programs protected by the obfuscator will hence contain the same data structures. 
Consequently, adversaries can try to locate the injected code manipulating the data structures with targeted techniques that result from a human or machine learning processes. Because the injected code fragments do not perform computations contributing to the original program, they are also vulnerable to attacks that identify semantically irrelevant code, such as generic deobfuscation.

Figures~\ref{fig:obfuscator}(b) and~\ref{fig:obfuscator}(c) illustrate our idea of \textit{supple} and \textit{flexible} obfuscators.
Where possible, such obfuscators inject protection code tailored to the application at hand, i.e., using data structures internal to the application itself instead of data structures implemented in the obfuscator.

The ideal case is illustrated by program 3 in Figure~\ref{fig:obfuscator}(b), in which a supple obfuscator uses the application's graph implementation to deploy a protection instead of its own hardcoded one. If a protection designed and implemented by security experts uses specific data structures, such as hash tables and graphs, and the unprotected application contains equivalents to these data structures, the protection reuses the application-specific implementation. 
In our mind, based on our experience, this will yield stronger, harder to defeat protection because the injected code is tuned to the original application's graph API, and hence will be not as easily spotted and learnable by adversaries.

In practice, however, the described supple obfuscation is ad hoc and hard to generalise. First, complex data structures used by a protection will not be available for reuse in all programs, such as program 4 in the figure. Secondly, some protections don't use any complex data structure at all.

We therefore present an alternative technique to reuse data structures. This alternative builds on the fact that, even in cases in which application data structures cannot be reused in the ideal way as described above, they can still be used to replace direct data dependencies with indirect ones. Instead of passing and storing data directly between ordinary variables and simple data structures within the protection code, that data can be stored temporarily in complex data structures reused from the original application. This is illustrated in Figure~\ref{fig:obfuscator}(c), in which we introduce the concept of a \textit{flexible} obfuscator.
In this example, the tradional implementation of an injected protection would store and pass some values in ordinary variables \textit{a}, \textit{b} and \textit{c}. Rather than injecting only the default implementation thereof, the obfuscator instead reuses the complex data structures already available in the original programs to store the data and to pass it around.
In general, anywhere in the protection code where a value is produced that will be assigned to a variable or anywhere an assignment to it occurs, the value can be stored or encoded in a complex data structure. Anywhere the variable is then later accessed, the value can be retrieved from the data structure.
Besides obfuscating the data flow, it is clear that the reuse of data structures helps in making the protection code look more similar to the application code. 

One option then is to simply store values in the complex data structures using native encodings. For example, integers can be stored in the keys or values of hash tables. Alternatively, values can also be encoded in features or shapes of the complex data structures. One example of this are the aliasing-based predicates we referred to earlier, but many other alternatives can be implemented as well: paths in a graph, the size of a hash table, or relations between values in a container are only some examples of how values can be encoded.

The stealth of the injected code heavily depends on the observable differences between that code and the application code. The more both look and behave alike, the more effort adversaries will have to invest in order to distinguish one from another.
There are several ways to make the protection code look similar.
First, data structures are preferably created and manipulated using the data structures' API functions already invoked in the original program. Typically, API invocations in the original program will target the data structure's public API. Sometimes, however, (part of) the private API will be invoked instead. This is the case, for example, in code that has been inlined by the compiler. In that case, that part of the private API can also be used.
Second, preferably API functions that are actually executed by the original application, at least for some inputs, are executed in the injected code.
This will hamper attacks that identify and expose the protection code and its related functionality by its (quasi-)invariant behaviour.
If then, e.g., the arguments and return values of the called API functions are quasi-invariant in the context of the protection code, the API functions will still appear as having variable arguments and return values because they are called with other values and return other values in the original application code. To identify the executed API functions, one can use several techniques, such as profiling and instruction tracing.
Third, similar data types and values as those in the original application are preferably used. Doing this will render the results of type-sensitive analysis useless. For example, if a program only stores string keys in hash tables, the injected code preferably also uses strings for that. Ideally, those strings are either already present in the original application statically, or analysis has shown that they are generated dynamically therein. A variety of tools and techniques exist to collect information on datatype usages and occurring values either statically, such as disassemblers and static analysis tools, or dynamically, such as debuggers and instrumentation tools.
Fourth and final, varying argument values for API functions called from within the protection code are preferably generated. This will lessen the amount of invariants present in said code, and, if those values depend on program input, the accuracy of the attacker's analyses will drop. One way to do this is to set an argument value within an existing if-then-else construction. In typical applications, the tested condition will be input-dependent, hence extending that input-dependency to the argument values.

\section{Flexible opaque predicates}
\label{sec:opaque}
To make the proposed technique more concrete and validate the general idea, we present a proof-of-concept design in which we inject bogus control flow into a program by deploying flexible OPs, i.e., OPs that build on application data structures and APIs already available in the original program.

Concretely, an OP consists of a computation resulting in a boolean value. The result of the computation is known by the obfuscator, but is hard to extract by reverse engineers without targeted techniques. Much research has been carried out on different forms of OPs.
Algebraic or number-theoretic predicates are the most basic form~\cite{arboit2002method,myles2006software}. Examples are the divisibility of some expression by a number, and well-known inequalities.
Concurrency-based predicates are also available, such as the ones proposed by Nagra et al., which exploit the difficulty of analysing the data and control flow of multi-threaded programs~\cite{nagra2004threading}.
And finally, some predicates are encoded in opaque invariants maintained by complex data structures. While the instantiated data structures are manipulated, and hence display variable behaviour, the injected manipulations maintain the invariants. The invariants in Collberg's approach relate to aliases between pointers to elements in manipulated graphs. The opaqueness of those invariants results from the fact that alias analysis is a hard static analysis problem~\cite{collberg2007dynamic}.

\begin{figure}[t]
  \centering
  \subfigure[One-way predicate]{\includegraphics[width=0.38\linewidth]{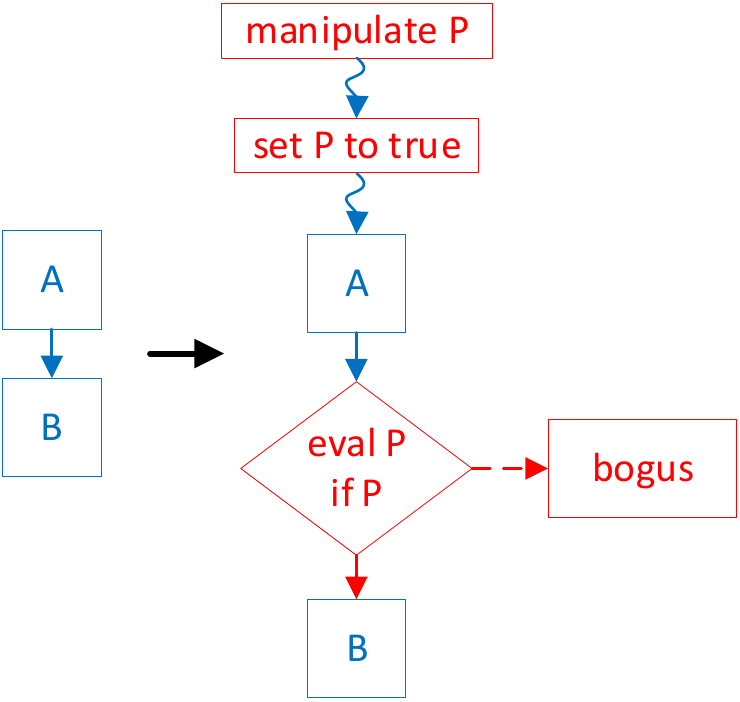}\label{fig:onewaypred}}
  \hspace{0.07\linewidth}
  \subfigure[Two-way predicate]{\includegraphics[width=0.45\linewidth]{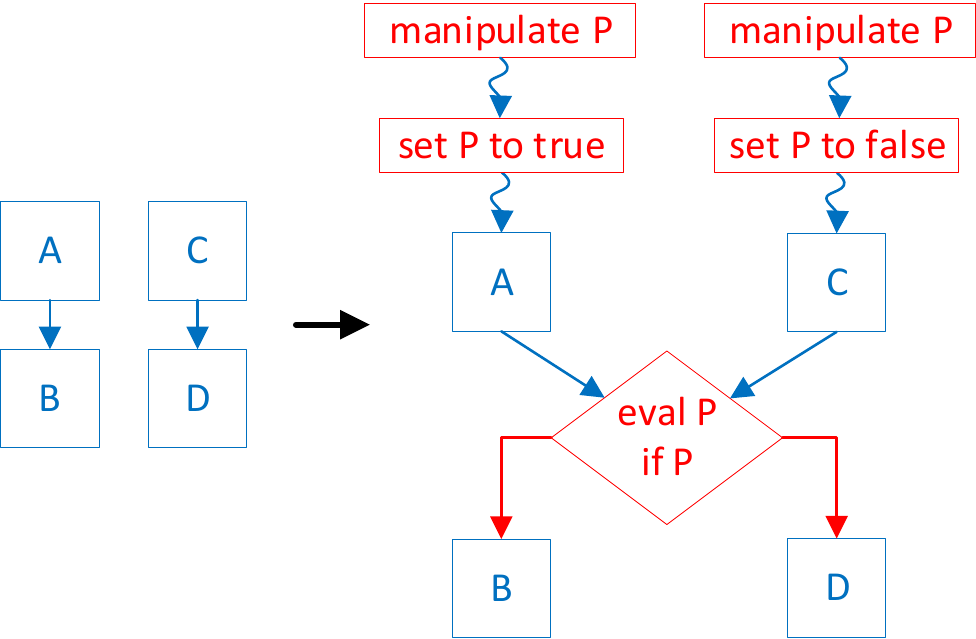}\label{fig:twowaypred}}
  \caption{Overview of predicate designs.}
  \label{fig:onetwoway}
  \Description{Fully described in the text.}
\end{figure}

In this research, we protect programs with one-way and two-way predicates.
A one-way OP is a predicate that invariantly evaluates to true or false. In that case, one of the control flow edges originating from the predicate is truly bogus: it will never be taken. The bogus control flow edge can hence be used to confuse an adversary and his tools by connecting it to an arbitrary location within the program~\cite{linn2003obfuscation}.
Figure~\ref{fig:onewaypred} illustrates the protection of a program with such a predicate, where part of an unprotected program is shown on the left and its protected version on the right. Specifically, some evaluation code is injected between blocks A and B to evaluate a predicate P, resulting in the introduction of a bogus control flow edge.
The evaluation code produces the OP value by means of a numeric computation, as is the case for algebraic predicates, or by means of a check of some invariant, as is the case with Collberg's alias-based predicates.

We consider two options to implement such OPs that check invariants on complex data structures. The first option involves \emph{flow-sensitive invariants}, i.e., invariants that hold at some program points but not at others. In the literature, these are also called contextual OPs. Figure~\ref{fig:onewaypred} illustrates one. To ensure that P evaluates to the required true (or false) at the conditional branch, some code is executed on the data structures that forces P to true (or false) on the path to the evaluation point. Prior to that last setting of P, the underlying data structures can have been manipulated such that a similar evaluation at an earlier point in the program would have resulted in another value.

The second option involves \emph{flow-insensitive invariants}, i.e., invariants that are maintained throughout the whole program execution. In that case, the underlying data structures are initialized at the start of the program to ensure that any check of the invariant at any program point will evaluate to the same value. Concretely, the ``set P to true'' computation in Figure~\ref{fig:onewaypred} is then not executed somewhere on the path to an evaluation point, but at the start of the program. The underlying data structures can then still be manipulated during the program's execution, which is useful to prevent attackers from observing the invariant all too easily, but only in a way that does not break the invariant. Collberg's alias-based predicates are an example of this: disjunct graphs and pointers to them are initialized upon program initialization, and the pointers traverse the graphs during the execution, but always keep pointing to different graphs and hence different nodes. 

Which of the two options, flow-sensitive or flow-insensitive, is the strongest is an open question. Research is needed to check which of the two is easiest to discover. One potential advantage of flow-sensitive invariants over flow-insensitive ones is that flow-sensitive ones can be complementary, in the sense that at some program points the data structures can be manipulated to force the predicate to true (i.e., set the predicate), while at other program points the data structures can be manipulated to force the predicate to false (i.e., reset the predicate). Local static analysis of the code fragment that evaluates the predicate at a single point in the program then does not immediately leak the predicate value to which other occurrences of the same or a similar fragment will lead. Instead a more global analysis will be needed that considers evaluation points and set/reset points.

Still, both alternatives for one-way OPs can easily be discovered with dynamic analysis, as each conditional branch on an OP is either always or never taken. It hence behaves quasi-invariantly, as is exploited in the generic deobfuscation attack by Yadegari et al.~\cite{yadegari15generic}.

By contrast, the result of a two-way OP evaluation varies at run time, and both outgoing edges of the conditional branch can be taken.  
One application of such predicates can be to randomly switch between two diversified variants of the same code fragment, depending on the computation's result~\cite{collbergbook}. Both variants need to be semantically equivalent, but can be implemented differently by means of code diversification techniques~\cite{hosseinzadeh2018diversification}. Such two-way predicates make the control flow and code appear more complex. Another application are dynamic OPs as proposed by Palsberg et al.~\cite{palsberg2000experience}. They build groups of opaquely related OPs, such as a group $\{P_1,P_2\}$ with the opaque relation $P_1 \Leftrightarrow \lnot P_2$. They transform straight-line code fragments such as $A ; B ; C ;$ into $\textrm{if} (P_1) \{A ; B;\} \textrm{else} \{A;\} ; \textrm{if} (P_2) \{B; C;\} \textrm{else} \{C;\}$, in which the then and else parts can be further diversified, and in which each realizable path implements the same semantics. Within each program execution, the predicates are all constant, but they can evaluate to different values in different runs, such that different paths can be executed in different runs of the obfuscated programs. Xu et al.\ generalized the techniques to code fragments involving control flow, such as loops, and to predicates of which the predicates' values can change within a single run~\cite{xu2016generalized}. The generalizations made their technique resilient against some attacks on (simple) dynamic OPs~\cite{ming2015loop}.

Van den Broeck et al. proposed a third application of two-way OPs, namely to connect unrelated code fragments from different functions, with the goal of hampering disassemblers that partitioning the code into components such as functions~\cite{van2020obfuscated}. They do this by merging control flow paths from different code contexts in different components, if possible in combination with outlining of equivalent fragments from those contexts. At the end of each merged path, a dispatcher diverts the control flow back to the original contexts. Two-way OPs are one of the possible forms for this dispatcher. This form of two-way predicates is illustrated in Figure~\ref{fig:twowaypred}, in which part of an unprotected program is shown on the left and its protected version on the right. The control flow of two paths A-B and C-D is merged after blocks A and C. At the merge point, some getter code is injected for the evaluation of predicate P, as well as some decision code to decide where the control flow should be redirected to. Control flow can be redirected to either B or D, depending on the predicate's value which has been set to complementary values on each merged code path before the execution of blocks A and C.
Van den Broeck et al.\ showed that, in combination with code layout randomization techniques, even relatively simple OPs can severely thwart the disassembly process of commonly used tools such as IDA Pro and Binary Ninja. They also showed that their technique is resilient to complex generic analyses such as value set analysis, generic deobfuscation and relatively simple pattern-matching techniques that search for uninterrupted OP computations. Advanced pattern-matching techniques are still possible, however, as the patterns of the injected protection code are hardcoded in the obfuscator.

Our proof of concept implementation applies the principles introduced in Section~\ref{sec:reuse} to one-way predicates used to inject bogus control flow like Collberg's alias-based OPs and to the two-way opaque dispatchers proposed by Van den Broeck et al.~\cite{van2020obfuscated}. For the one-way predicates, we support flow-sensitive as well as flow-insensitive invariants in complex data structures. For the two-way OPs, we of course only support flow-sensitive ones: by definition, the predicate needs to be set on one incoming path, and reset on the other incoming path. In such cases, the term invariant can be somewhat misleading, as the predicate evaluation in the dispatcher does not invariantly evaluate to either true or false. The term is still correct, however, with the invariants being of the form ``if the dispatcher is reached through block A, then P will evaluate invariantly to true.'' We will refer to this as \emph{context-sensitive invariants} in the remainder of the paper. Clearly, both flow-sensitive one-way OPs and context-sensitive two-way OPs are forms of contextual OPs. 

Listing~\ref{lst:protected} provides a concrete example of our approach for one-way, flow-sensitive predicates deployed on a graph-processing program, in which the evaluated invariant is that a graph contains at least two non-adjacent nodes. In the code listing, the original application code is marked black and the injected protection code is marked red.
The injected one-way OP is evaluated at line~\ref{pseudo:test}. The predicate's evaluation depends on the values of variables \textit{K} and \textit{L}, which are initially set at lines~\ref{pseudo:K} and~\ref{pseudo:L}. Between this setter code and the predicate's evaluation code, a call to function \textit{C} is performed in which the predicate's value is optionally manipulated at lines~\ref{pseudo:KK} and~\ref{pseudo:LL}, introducing an input dependency on \textit{param}. As is clear from its evaluation at line~\ref{pseudo:test}, the predicate itself is defined on an instance of the application-specific \textit{Graph} implementation. Specifically, it is evaluated on the original application's instance \textit{G}, but an additional instance \textit{X} is required to implement the predicate. That instance is created at line~\ref{pseudo:copy} as a deep copy of \textit{G}.
Note that the predicate and its setter code are implemented with functionality that is already present and invoked in the original application. The invoked functions can hence not serve as direct indicators of a protection being present. Moreover, they will hence not show invariant behaviour internally, which will prevent that their invocations can be omitted as part of trace simplification based on quasi-invariant behaviour.

The same listing can also be used to illustrate our approach on two-way, context-sensitive OPs. To this end, consider the additional code in Listing~\ref{lst:protectedtwoway}, assuming that the function \textit{E} is called upon by the original application. In that function, variables \textit{K} and \textit{L} are set in order to make the predicate evaluate to a value that is complementary to the one set in Listing~\ref{lst:protected}. The predicate's evaluation is initiated by the \textit{goto} at line~\ref{pseudo:gotoX}\footnote{Interprocedural control flow redirection by means of labels and \textit{goto} statements is not allowed in high-level programming languages such as C, but it is allowed at lower levels such as assembly code.}, after which the control flow will return to the \textit{bogus} label at line~\ref{pseudo:Xreturn}. The check at line~\ref{pseudo:test} will hence display variable behaviour, which prevents it from being eliminated by the trace simplification of Yadegari et al~\cite{yadegari2015symbolic}.

\begin{lstlisting}[frame=single, escapechar=|, breaklines, label={lst:protected}, caption={Example program (black) with protected code (\textcolor{red}{red}).}, float, morekeywords={function, string, int, for, in, return, if, else, bool, new, label, goto}]
Graph *G[*@, *X@*]|\label{pseudo:instance}|
[*@Node *K, *L@*]|\label{pseudo:nodes}|

function A(string file, bool param) {|\label{pseudo:Abegin}|
  [*@int temp1 =@*] B(file)|\label{pseudo:callB}|
  [*@K = X.add_node(temp1)@*]|\label{pseudo:K}|
  [*@L = G.get_node(0)@*]|\label{pseudo:L}|
  C(G.get_node(0), param)|\label{pseudo:callC}|
[*@label location_get:@*]|\label{pseudo:label}|
  [*@if (!G.adjacent(K, L))@*]|\label{pseudo:test}|
    D()
  [*@else@*] [*@goto bogus@*]|\label{pseudo:goto}|
}|\label{pseudo:Aend}|

function B(string filename) {|\label{pseudo:Bbegin}|
  G = new Graph()|\label{pseudo:constructor}|
  int edge_count = 0
  for [from, to] in File(filename)
    Node *I, *J
    I = G.add_node(from)
    J = G.add_node(to)
    G.add_edge(I, J)|\label{pseudo:fill}|
    edge_count += 1
  [*@X = G.deep_copy()@*]|\label{pseudo:copy}|
  return edge_count
}|\label{pseudo:Bend}|

function C(Node *root, bool param) {|\label{pseudo:Cbegin}|
  for n in G.reachable_from(root)
    for m in G.get_neighbours(n)
      if (... && param)|\label{pseudo:paramtest}|
        G.del_edge(n, m)
        ...
        [*@X.add_edge(K, K)@*]|\label{pseudo:KK}|
      else
        [*@L = X.get_node(m.index)@*]|\label{pseudo:LL}|
        ...
        G.add_edge(m, n)
  int temp2 = G.do_something()|\label{pseudo:dosomething}|
  G.do_something_else(temp2)|\label{pseudo:dosomethingelse}|
}|\label{pseudo:Cend}|

function D() {|\label{pseudo:Dbegin}|
  for n in G.nodes
    if (G.adjacent(n, n))|\label{pseudo:adjacent}|
      G.del_edge(n, n)|\label{pseudo:deledge}|
}|\label{pseudo:Dend}|
\end{lstlisting}

\begin{lstlisting}[firstnumber=51, frame=single, escapechar=|, breaklines, label={lst:protectedtwoway}, caption={Example program (black) with protected code (\textcolor{red}{red}).}, float, morekeywords={function, string, int, for, in, return, if, else, bool, new, label, goto}]
function E() {
  [*@int temp = ...@*]
  [*@K = X.get_node(temp)@*]|\label{pseudo:XK}|
  [*@L = X.get_neighbours(K).get(0)@*]|\label{pseudo:XL}|
  [*@goto location_get@*]|\label{pseudo:gotoX}|
[*@label bogus:@*]
   ...|\label{pseudo:Xreturn}|
}
\end{lstlisting}

In the example, only calls to public API functions have been injected. However, users of a flexible obfuscator are not limited to such public APIs. Because public API functions, in particular wrappers, getters and setters, are often inlined by the compiler for performance reasons, it may well be that compiled programs contain calls to private API functions and direct accesses to data structures. In such cases, the injected code may of course also contain similar calls and data accesses.

In the example, an invariant based on the adjacency of graph nodes was implemented because it can be evaluated and maintained by calling API functions and using data structures already present in the original program. Ideally, the obfuscator supports as many different kinds of invariants on different kinds of data structures as its user can imagine. The user of the obfuscator must therefore have as much freedom as possible to specify which data structures are to be used, how they need to be initialized and manipulated, and how the invariants will be checked. To that end, the user needs to be able to describe the code chunks to be injected, including calls to existing functions, preferably on an abstract level, and the obfuscator needs to be able to inject them. Obviously, the obfuscator must also be able to inject additional control flow, because bogus and complicated control flow is the actual goal of the obfuscation.

Concerning the manipulation of control flow, our flexible obfuscator does not differ from a fixed obfuscator, which is why we don't elaborate on this aspect any further. Instead, we will focus on the meta-API with which the user of the flexible obfuscator can provide the necessary inputs.

\section{A meta-API for flexible obfuscators}
In this section, we will describe the meta-API with which the flexible obfuscator's user can provide the required prescriptions.

\subsection{Support for calling the application's API}
To correctly inject function calls, the obfuscator needs to know the signature of each callee: the number of arguments, the data type of each argument and, if relevant, the return value's data type.

If the flexible obfuscator is implemented as a source-to-source rewriter, or if it is integrated into a compiler, then the signatures of the required API functions and the definitions of the involved data structures are available by design. As those functions and data structures are called and used by the original application, the compiler needs to know their signatures and definitions, or it would not be able to build the original program.

By contrast, binary rewriting obfuscators operate on already compiled code instead. Source-level information might then still be available, in the form of source code, or in debug information (e.g., DWARF~\cite{dwarfspec}), or run-time type information (RTTI) available in the object files that were generated by the compiler.\footnote{The debug information only needs to be stripped from the binary after it has been rewritten.} It might be, however, that no source-level information is available, as when pre-compiled third-party C components are linked into a program and the third parties do not want to share source-level information as rich as debug information. In that case, it is still possible to provide the strictly necessary information in some custom format.
The binary rewriter with which we implemented our prototype does not have full support for parsing DWARF debug information nor does it support parsing RTTI. Hence, we opted for a custom XML format with which the user provides the signature descriptions. The design of a custom format is more of an engineering issue rather than a research topic however, which is why we don't elaborate on this any further.

\subsection{Support for injecting flexible opaque predicates}
\label{sec:inject_support}
To support the injection of protections, in our case control flow obfuscations based on flow-insensitive, flow-sensitive, and context-sensitive OPs based on complex data structures, the obfuscator needs to inject code fragments of varying complexity. As shown in the example listing, they can include programming idioms such as statically allocated data, local variables, dynamically allocated data, the already discussed function calls, etc.

Code fragments need to be injected for the following actions:
\begin{itemize}
\item \textbf{Initialize global data structures}: As the protections will make use of complex data structures, those will need to be set up somehow. This can be done statically, dynamically, or with a combination thereof. In dynamic approaches, this can either be done by means of code fragments injected into the main application code, or by means of initializer functions.
\item \textbf{Evaluate the predicate and branch off it}: Wherever bogus control flow or two-way dispatchers will be inserted, the data structures need to be queried to obtain the predicate value. 
\item \textbf{Set and reset predicates}: In the case of flow-sensitive or context-sensitive invariants, the obfuscator needs to inject fragments setting and resetting the predicates by manipulating the underlying data structures on the paths leading to predicate evaluations.
\item \textbf{Orthogonal data structure manipulations}: Whatever form of invariant is used, we can make it harder for attackers to observe them and to analyze the predicates' evaluations by inserting additional manipulations of the underlying data structures throughout the program. In the case of flow-insensitive invariants, these manipulations can alter the underlying data structures but not in a way that would break the invariant. With the other invariants, manipulations can also set or reset the predicate, and they can even do so in unpredictable ways. In other words, it is possible to prescribe manipulations of which the obfuscator does not know what the impact on the predicate will be. Following any program point where such manipulations are injected, another injected manipulation will always predictably set or reset (as needed) the predicate value before an evaluation of it is reached.
\end{itemize}

It is the user of this flexible obfuscator that needs to provide these code fragments. In our research, we did not yet explore what would be the best (pseudo-)code syntax for prescribing them. While we consider it an interesting research question, syntax is out of the scope of this paper. So is the engineering that is necessary to support specific syntaxes. Instead we will here focus on semantics, with two objectives. First, we aim to give the user maximal flexibility to explore any forms of invariants on any kind of data structure they can imagine. Secondly, we aim for letting the user express his imagination at an abstract level. We do so for the user's convenience and because this will give the obfuscator more freedom to implement the fragments in stealthy and diversified ways. 

In light of these goals, we opt to let the user provide pseudo-code fragments augmented with pre-conditions and post-conditions, plus some very basic auxiliary information. With post-conditions, the user can describe the effect of each code fragment on the predicates, i.e., whether a predicate is set, reset, or becomes unknown. With pre-conditions, the user can prescribe the constraints on the inputs of the code fragments that need to be met in order to achieve the given post-condition.   

For the example in Listing~\ref{lst:protected}, the user first needs to specify the global variables that need to be defined, being \textit{X}, \textit{K}, and \textit{L}. On them, a predicate \textit{initialised()} is defined that is set to false at the program entry point. This predicate can be used in pre- and post-conditions to indicate that initialization fragments need to be injected. Table~\ref{tab:init} shows an example. The post-condition states that \textit{X} will be initialized when this fragment has been executed. If an initialization routine has no pre-conditions, it can be inserted anywhere in the program. The obfuscator can then use control flow analyses such as interprocedural dominator analysis~\cite{de2007practical} to determine the program points that will precede any insertion of actual OP manipulations or evaluations. At least the program entry point is one of them. In the example, however, the precondition specifies that the initialization routine can only be injected at program points where \textit{G} has been initialized, i.e., when line 27 of the example is reached. Obviously, a flexible obfuscator cannot determine that point without assistance from the user. So for such initializers, we expect the user to provide a description of the point where the initialization can be inserted. This can easily be done by means of pragmas in C or C++ source code, or by an external description in terms of source line numbers or locations in binary code sections. Requiring such descriptions of relevant program points does not impose a large additional burden on the users of the protection tool, as they already have to indicate the program regions on which they want to deploy software protections anyway. 

{\small
\begin{table}[t!]
  \begin{minipage}{0.49\linewidth}
    \centering
    \captionof{table}{Initialisation fragment for $\mathcal{P}$.}
    \label{tab:init}
    \vspace{-1em}
    \begin{tabular}{|l|l|}
      \hline
      \textbf{Pre-cond}  & \begin{tabular}[c]{@{}l@{}}
        initialised(G) = true \\
      \end{tabular} \\ \hline
      \textbf{Post-cond}  &
      initialised(X) = true
      \\ \hline
      \textbf{Fragment} & \begin{tabular}[c]{@{}l@{}}
        X = G.deep\_copy()
      \end{tabular} \\ \hline
    \end{tabular}
    \vspace{1em}
    \captionof{table}{Setter fragment for $\mathcal{P}$.}
    \label{tab:set}
    \vspace{-1em}
    \begin{tabular}[t]{|l|l|}
      \hline
      \textbf{Inputs}  &
      i
      \\ \hline
      \textbf{Pre-cond}  & \begin{tabular}[c]{@{}l@{}}
        initialised(X) = true \\
        type(i) = integer
      \end{tabular} \\ \hline
      \textbf{Post-cond}  & \begin{tabular}[c]{@{}l@{}}
        $\mathcal{P}$ = true \\
        $\mathcal{Q}$ = unknown\\
        initialised(K) = true \\
        initialised(L) = true
      \end{tabular} \\ \hline
      \textbf{Fragment} & \begin{tabular}[c]{@{}l@{}}
        K = X.add\_node(i)\\
        L = G.get\_node(0)
      \end{tabular} \\ \hline
    \end{tabular}
  \end{minipage}
  \begin{minipage}{0.49\linewidth}
    \centering
    \captionof{table}{Resetter fragment for $\mathcal{P}$.}
    \label{tab:reset}
    \vspace{-1em}
    \begin{tabular}{|l|l|}
      \hline
      \textbf{Inputs}  &
      i
      \\ \hline
      \textbf{Pre-cond}  & \begin{tabular}[c]{@{}l@{}}
        initialised(X) = true \\
        type(i) = integer\\
        value(i) $\in$ [0:10:2]
      \end{tabular} \\ \hline
      \textbf{Post-cond}  & \begin{tabular}[c]{@{}l@{}}
        $\mathcal{P}$ = false \\
        $\mathcal{Q}$ = unknown\\
        initialised(K) = true \\
        initialised(L) = true
      \end{tabular} \\ \hline
      \textbf{Fragment} & \begin{tabular}[c]{@{}l@{}}
        K = X.get\_node(i)\\
        L = X.get\_neighbours(K).get(0)
      \end{tabular} \\ \hline
    \end{tabular}
    \vspace{1em}
    \captionof{table}{Evaluation fragment for $\mathcal{P}$.}
    \label{tab:eval}
    \vspace{-1em}
    \begin{tabular}{|l|l|}
      \hline
      \textbf{Pre-cond}  & \begin{tabular}[c]{@{}l@{}}
        initialised(K) = true \\
        initialised(L) = true
      \end{tabular} \\ \hline
      \textbf{Result}  & \begin{tabular}[c]{@{}l@{}}
        $\mathcal{P}$
      \end{tabular} \\ \hline
      \textbf{Fragment} & \begin{tabular}[c]{@{}l@{}}
        !G.adjacent(K, L)
      \end{tabular} \\ \hline
    \end{tabular}
  \end{minipage}
  \vspace{-1em}
\end{table}
}

Tables~\ref{tab:set} and~\ref{tab:reset} show the prescriptions of the set and reset fragments used in the example listings. When those fragments are injected, the obfuscator has to insert the necessary glue code that provides inputs to those fragments for the mentioned input variables. The pre-conditions include requirements that those inputs need to be of a certain type and (in Table~\ref{tab:reset}) that their values need to meet certain conditions (in this case being an even number in the range $[0,10]$). Additional pre-conditions specify that the accessed global variables need to have been initialised at the point where the fragments are injected. We should note that if the user knows that the necessary initialization will have been done before any of the code regions in which the obfuscations are injected can be reached, he can simply omit those pre-conditions. The tool then does not need to deploy complex control flow analysis that potentially might lack the necessary precision. The post-conditions of the setter and resetter fragments indicate that predicate $\mathcal{P}$ will be have been set to \textit{true} or \textit{false}, respectively. The obfuscator assumes that at the program entry point, $\mathcal{P} = unknown$. The predicates are hence treated as tri-state predicates with the possible values \textit{false}, \textit{true}, and \textit{unknown}. 

Table~\ref{tab:eval} shows the fragment used to evaluate the predicate. For evaluation functions, we prescribe the result value, which in this case is simply $\mathcal{P}$. Notice the lack of pre-conditions on $\mathcal{P}$ for this evaluation function. It is up to the flexible obfuscator to combine setters and resetters with evaluation fragments in the correct way to construct valid one-way or two-way OPs.

In the example of Listing~\ref{lst:protected} only one predicate is stored in the re-used data structure, corresponding to $\mathcal{P}$ in the tables, but nothing prevents the user from defining multiple ones. In fact, it is highly advised to do so, because it will result in a wider range of data manipulations of the data and hence in stealthier and more resilient deployment. Tables~\ref{tab:setQ} and~\ref{tab:evalQ} illustrate how a second predicate $\mathcal{Q}$ can be defined. In this case, $\mathcal{Q}$ is not orthogonal to $\mathcal{P}$, as indicated by $\mathcal{P}$ becoming unknown according to the post-conditions.

Table~\ref{tab:setQ} also illustrates how prescribing multiple combinations of pre- and post-conditions on the same code fragment can be used to set or reset $\mathcal{Q}$ with the same fragment depending on the pre-conditions of the input variables. In this case,  $\mathcal{Q}$ is set if \textit{i} is even, and reset of it is odd.

{\small
\begin{table}[t]
  \begin{minipage}{0.49\linewidth}
    \centering
    \caption{(Re)setter fragment for $\mathcal{Q}$.}
    \label{tab:setQ}
    \vspace{-1em}
    \begin{tabular}{|l|l|}
      \hline
      \textbf{Inputs}  &
      i
      \\ \hline
      \textbf{Pre-cond 1}  & \begin{tabular}[c]{@{}l@{}}
        initialised(X) = true \\
        type(i) = integer\\
        value(i) $\in$ [0:10:2]
      \end{tabular} \\ \hline
      \textbf{Post-cond 1}  & \begin{tabular}[c]{@{}l@{}}
        initialised(M) = true \\
        $\mathcal{P}$ = unknown\\
        $\mathcal{Q}$ = true 
      \end{tabular} \\ \hline
      \textbf{Pre-cond 2}  & \begin{tabular}[c]{@{}l@{}}
        initialised(X) = true \\
        type(i) = integer\\
        value(i) $\in$ [1:10:2]
      \end{tabular} \\ \hline
      \textbf{Post-cond 2}  & \begin{tabular}[c]{@{}l@{}}
        initialised(M) = true \\
        $\mathcal{P}$ = unknown\\
        $\mathcal{Q}$ = false 
      \end{tabular} \\ \hline
      \textbf{Fragment} & \begin{tabular}[c]{@{}l@{}}
        M = X.get\_node(i)\\
        if (i \% 2)\\
        \hspace{0.5em}X.add\_edge(M, M)\\
        else\\
        \hspace{0.5em}X.remove\_outgoing\_edges(M)\\
      \end{tabular} \\ \hline
    \end{tabular}
  \end{minipage}
  \begin{minipage}{0.49\linewidth}
    \centering
    \caption{Evaluation fragment for $\mathcal{Q}$.}
    \label{tab:evalQ}
    \vspace{-1em}
    \begin{tabular}{|l|l|}
      \hline
      \textbf{Pre-cond}  & \begin{tabular}[c]{@{}l@{}}
        initialised(X) = true \\
        initialised(M) = true 
      \end{tabular} \\ \hline
      \textbf{Result}  & \begin{tabular}[c]{@{}l@{}}
        $\mathcal{Q}$
      \end{tabular} \\ \hline
      \textbf{Fragment} & \begin{tabular}[c]{@{}l@{}}
        X.get\_neighbours(M).size() > 0
      \end{tabular} \\ \hline
    \end{tabular}
    \vspace{1em}
    \caption{Initialisation code fragment for $\mathcal{R}$.}
    \label{tab:htinit}
    \vspace{-1em}
    \begin{tabular}{|l|l|}
      \hline
      \textbf{Pre-cond}  & \begin{tabular}[c]{@{}l@{}}
        initialised(H) = true \\
      \end{tabular} \\ \hline
      \textbf{Post-cond}  &
        initialised(X) = true
      \\ \hline
      \textbf{Fragment} & \begin{tabular}[c]{@{}l@{}}
        X = H.deep\_copy()
      \end{tabular} \\ \hline
    \end{tabular}
    \vspace{1em}
    \caption{Setter code fragment for $\mathcal{R}$.}
    \label{tab:htsettrue}
    \vspace{-1em}
    \begin{tabular}{|l|l|}
      \hline
      \textbf{Inputs}  &
        key, value
      \\ \hline
      \textbf{Pre-cond}  & \begin{tabular}[c]{@{}l@{}}
        initialised(X) = true \\
        type(key) = string \\
        type(value) = integer
      \end{tabular} \\ \hline
      \textbf{Post-cond}  & \begin{tabular}[c]{@{}l@{}}
        $\mathcal{R}$ = true\\
        \_key = key
      \end{tabular} \\ \hline
      \textbf{Fragment} & \begin{tabular}[c]{@{}l@{}}
        X.insert(key, value)
      \end{tabular} \\ \hline
    \end{tabular}
  \end{minipage}
\end{table}
}

Finally, we need to discuss one more feature of the meta-API. Tables~\ref{tab:htinit} to~\ref{tab:hteval} prescribe the code fragments for another predicate $\mathcal{R}$ that is defined on a hash table and that depends on whether or not some key is present. Obviously, the key being checked for the evaluation of the predicate should be the same as the key used for setting or resetting the predicate. We do not want to hardcode the keys to be used, however, so we should be able to prescribe this constraint. We can do so by means of additional variables that do not show up in the injected code, but that the obfuscator uses internally (similar to the \textit{initialised()} predicates) to ensure that only valid combinations of setters, resetters, and evaluations are injected. In Tables~\ref{tab:htinit} to~\ref{tab:hteval}, the internal variable is \_key. When injecting a setter or resetter, the obfuscator can freely choose any string as key, as allowed by the pre-conditions. While no code will be injected to do so, the obfuscator considers \_key to be set to key at the end of the injected fragment because of the specified post-condition on \_key. For the evaluation fragment, the obfuscator can then not choose the input key anymore, as the pre-condition states it is bound to \_key. 

{\small
\begin{table}[t]
  \begin{minipage}{0.49\linewidth}
    \centering
    \caption{Resetter code fragment for $\mathcal{R}$.}
    \label{tab:htsetfalse}
    \vspace{-1em}
  \begin{tabular}{|l|l|}
    \hline
    \textbf{Inputs}  &
      key
    \\ \hline
    \textbf{Pre-cond}  & \begin{tabular}[c]{@{}l@{}}
      initialised(X) = true \\
      type(key) = string
    \end{tabular} \\ \hline
    \textbf{Post-cond}  & \begin{tabular}[c]{@{}l@{}}
      $\mathcal{R}$ = false \\
      \_key = key
    \end{tabular} \\ \hline
    \textbf{Fragment} & \begin{tabular}[c]{@{}l@{}}
      X.remove(key)
    \end{tabular} \\ \hline
  \end{tabular}
  \end{minipage}
    \begin{minipage}{0.49\linewidth}
      \centering
    \caption{Evaluation code fragment for $\mathcal{R}$.}
    \label{tab:hteval}
    \vspace{-1em}
    \begin{tabular}{|l|l|}
      \hline
      \textbf{Inputs}  &
        key
      \\ \hline
      \textbf{Pre-cond}  & \begin{tabular}[c]{@{}l@{}}
        initialised(X) = true \\
        key = \_key
      \end{tabular} \\ \hline
      \textbf{Result}  & \begin{tabular}[c]{@{}l@{}}
        $\mathcal{R}$
      \end{tabular} \\ \hline
      \textbf{Fragment} & \begin{tabular}[c]{@{}l@{}}
        X.get(key)!=NULL
      \end{tabular} \\ \hline
    \end{tabular}
  \end{minipage}
  \vspace{-1em}
\end{table}
}

\subsection{Stealthy fragment injection}
\label{sec:stealthy_injection}
The fragments to manipulate the data structures and the encoded predicates as presented in the previous sections have inputs on which some constraints are placed by means of pre-conditions. Within those constraints, the flexible obfuscator has some freedom to choose where the inputs will come from at run time, and to generate and inject corresponding glue code around the injected fragments. This choice has an impact on stealth and on the ease with which attackers might comprehend what is going on in the code, and hence where and what the OPs are. At least two aspects matter: variability of the thus obtained inputs, and similarity to types and data (patterns) already occurring in the original program. The latter is particularly important for inputs of the fragments that are passed to API functions invoked on the data structures. If those stand out too much and seem out of place, it will be easy for attackers to spot one or more ``weird'' API calls, hypothesize that those correspond to obfuscations, and write custom scripts to automate the identification of all similarly standing out API calls, i.e., to generalize the findings of the attacker. Variability of input values between different occurrences of the same fragment (i.e., when a fragment is injected at multiple program points) is important to block this generalization.\footnote{Obviously, it is also important that the binary code generated for the injected fragments is not identical everywhere. For the time being, we assume software diversification and global compiler optimizations (i.e., optimizations that let the form of a particular generated code fragment depend on the surrounding code) can be used to generate enough syntactic diversity in that respect.}

Run-time variability of the inputs of a specific occurrence of a fragment is equally important, but for another reason: to make precise automated analyses and deobfuscation as well as human comprehension harder. Yadegari's generic deobfuscation technique relies specifically on data used in obfuscations being quasi-invariant, i.e., the same values reoccur throughout the execution of a program on some input. Fuzzing also performs much better if inputs to code fragments can exhibit less variation and less code paths are hence triggered. So does symbolic execution when certain (input) values can be shown or assumed to be constant. Even when humans observe likely invariants, such as when inputs to certain functions are always the same, they also simplify their mental model of the code based on the assumption that, until proven otherwise, the likely invariant is an actual invariant. So a lack of run-time variability of inputs to any instance of an injected fragment will ease both automated analysis and human comprehension by the attacker. 

The obfuscator has various options to find input values in the original program. For primitive types, local and global variables are options in source-to-source or compiler-based obfuscators. In binary rewriters, local data available in registers or in a stack frame are options, as well as global data in statically allocated (read-only) data sections, including string literals.  All of those data might be used directly or, if necessary, they might first be manipulated to meet pre-conditions such as the fact that a number needs to be odd or even, positive or negative, be in or out of some range, etc. Alternatively, the obfuscator can create primitive inputs from scratch, be it constant inputs or variable ones. Constant values can be obfuscated with many obfuscation techniques~\cite{tiella2017automatic,collberg1997taxonomy,zhou2007information}. Note that even in cases where no debug information is available, binary rewriters can differentiate among the available registers to select inputs from. For example, our proof-of-concept flexible obfuscator that we used for the empirical evaluation in Section~\ref{sec:evaluation} performs simple data flow analyses to determine whether data in a register is used as an address in memory accesses, whether it is a constant value or not, whether it can be a zero or not, a copy of some other register, etc. 

To decide how much variability is wanted, the original program can first be traced, after which statistics can be computed on the values passed to API functions. In the case source code is available, the source code can also be analyzed statically to find out which types of arguments are passed to API functions, but that information can also be retrieved by means of traces and debug information. For example, in the case of a C++ hash table that accepts any type of key and value objects (as long as the key objects provide a hash method), a trace or static analysis can be used to find out which concrete types are actually passed in the hash tables. To find inputs that maximally provide the wanted variability, the original program can first be traced, such that the variability of values produced and stored during the program can be assessed. Moreover, a taint-tracking analysis can be performed to identify the values that actually depend on input to the program. This will help in defeating attacks based on symbolic execution and generic deobfuscation. 

The mentioned pre-pass techniques to collect information enable the flexible obfuscator to some extent to make decisions on its own to find the most appropriate inputs to pass to the injected code fragments. The user can also steer the obfuscator, however, by means of the pre-conditions and post-conditions. For example, the \textit{type()} pre-conditions of the fragments in Tables~\ref{tab:htsettrue} and~\ref{tab:htsetfalse} already specify that the input key needs to be of type string. Additional pre-conditions can easily be formulated for specifying whether inputs are (preferably or mandatory) to be based on data that is pre-existing or not, global or local, constant or read-only, etc. Pre-conditions can also be formulated on the variability over different occurrences of the injected fragment, e.g., whether the obfuscator should strive for different values being passed at different instances or not. 

In addition, the user can provide additional code fragments that the obfuscator can inject not to set or reset a predicate, but to generate appropriate inputs for predicate setters and resetters. For example, the user might provide a set of custom fragments with a pre-condition \textit{type(i) = integer} and a post-condition \textit{value(i) $\in$ [1:10:2]}. With such fragments, the user can extend the limited, predetermined set of computations that the flexible obfuscator knows to inject for generating even numbers in a range. By supporting user-provided fragments to compute inputs to setters and resetters, the attacker's learnability of that part of the flexible protections can be reduced. Moreover, users can then choose and tune the provided fragments to achieve the desired variability of input values to inject setters and resetters, based on their knowledge of the behavior of the original application they want to protect.

\subsection{Generalization}
So far, we discussed how the meta-API supports flow-insensitive and contextual one-way and two-way OPs. As it is, the meta-API can also be used for more advanced types of OPs, such as dynamic OPs~\cite{palsberg2000experience}, generalized dynamic OPs~\cite{xu2016generalized}, bi-OPs~\cite{xu18}, range dividers~\cite{banescu2016code}, etc. To support those, the necessary decision logic and transformations need to be implemented in the flexible obfuscation tool, and the necessary relations between set, reset, and evaluation fragments need to be expressed with the appropriate pre- and post-conditions. Supporting the more advanced types of OPs might require some more engineering, but no fundamentally more complex concepts.

Furthermore, the injection of OPs is not the only application supported by the meta-API. Another interesting application is that of implicit information flow, which are also known as covert channels~\cite{stephens2018probabilistic,obf_anti_symbolicdeobf}. With those, direct data dependencies (i.e., assignments and copy operations) are rerouted through external system state to hamper taint-analysis and other data flow analyses that only consider internal program state. The meta-API allows users of a flexible obfuscator to introduce any implicit information flow they can imagine to thwart (future) analysis toolboxes that include countermeasures against fixed, known forms of implicit information flow. Actually, the meta-API can also be used to reroute direct data dependencies through explicit, but more complex data flow that reuses data structures and API functions already available in the program. As discussed in Section~\ref{sec:reuse}, this can be useful to diversify linked-in protection code and to tailer it to the protected program, thus making it harder for attackers to identify the linked-in protections.  

Finally, the meta-API can be used to avoid the re-use of fixed and hence detectable opaque constant generators, i.e., constant data values such as strings that are not hard-coded as data in a program but that are replaced by procedural code that generates their value dynamically (e.g.,~\cite{tiella2017automatic}). 

In summary, we conjecture that the generic nature of the meta-API and its use of pre- and postconditions enable support for a much wider range of flexible obfuscations than the OPs we experiment with in this paper. 

\section{Security Analysis - The Case for Flexible Software Obfuscation}
\label{sec:evaluation}
As for the potency of the proposed flexible OPs, this paper does not make new claims. The bogus control flow paths and other complications of code and data flow that can result from those paths are independent of their flexible or fixed nature, so flexible OPs give us the same potency as already evaluated in the extensive literature on OPs as discussed in Section~\ref{sec:opaque}. As for the hint in the introduction that mixing original program functionality and protection functionality by reusing the former for the latter might potentially improve the potency of protections, we have no empirical evidence yet. Lacking any new claims regarding potency, we then do not evaluate potency, and instead focus on resilience and stealth.

In the context of flexible OPs, resilience and stealth are almost synonymous: Once attackers identify a predicate as an OP of a certain type (always true, always false, two-way, ...), the hard work is done. How they then use that gained knowledge to undo the obfuscation or to bypass it or in some other way depends on their modus operandi and the specific tools they use, but is independent of whether the OP was a flexible or a fixed one. The security analysis in this section therefore focuses jointly on the stealth and resilience of flexible OPs, i.e., the capability of remaining undetected and not being deobfuscated. As a starting point, we consider the work by Zobernig et al., which surveyed eight common attacks on one-way OPs~\cite{zobernig19}. We discuss those eight attacks, and extensions thereof on two-way OPs, plus additional attacks and attack steps. 

Conceptually, most attacks begin with some sort of program inspection to identify the conditional branches steerd by potential OPs. Program slices can then optionally be constructed for those branches, in which attacks can be performed. Slices can be constructed statically or dynamically, and an attacker can try to simplify dynamic slices, which correspond to parts of execution traces, in a number of ways~\cite{yadegari15generic,blazytko2017syntia}.

\subsection{Brute Force Search}
In a brute force search attack, the attacker executes the extracted slice on all possible input value combinations to check whether the possible values of the predicate. While this can work for simple numerical (e.g., algebraic) opaque predicates of which the slices only take (narrow enough) numerical values as inputs, we conjecture that generalizing such attacks to well designed flexible OPs of which the values depend on complex datastructures' shapes and values, is impossible. Deciding what slice to consider is a hard problem in the first place, and for any slice of practical length, the space of possible inputs to consider (i.e., all possible states of the involved data structures), will be impractically large. In fact, the attacker has no idea what the real space of potential inputs looks like, as he is starting from a binary that lacks high-level information such as type information and as he has no a priori knowledge about the invariants maintained by the data structures throughout the program's execution. In an attack on the example of Listing~\ref{lst:protected}, for example, if the attacked considers the slice within function \texttt{A}, he doesn't know to what data structures \texttt{X} and \texttt{G} point, whether they point to data structures in disjount or overlapping memory regions, what values can already be stored in the pointer chains starting at \texttt{X} and \texttt{G}, etc. The attacker therefore has no basis to decide which potential memory states should be considered as potential inputs. Not being limited to states that can actually occur in the program, it is not hard to come up with memory states that would make the OP on line 10 evaluate in either direction. Not knowing the space of potential inputs is a sufficiently big problem to make this form of attack fail. If the considered slices are large enough (e.g., starting at the program entry point), the space might be known at the conceptual level. But for any non-trivial program, the space will then be too big to search exhaustively. 

\subsection{Evaluate at Zero}
Zobernig et al.\ proposed this search-space pruning heuristic themselves, based on their observation that in typical programs most integer comparisons with constants compare with zero. They therefore propose to discard a predicate as a potential OP if it evaluates to true for the zero input. Obviously, this attack can only work for numerical OPs, not for OPs based on complex data structures. 

\subsection{Probabilistic Checking}
In these attacks, not all the potential inputs are fed to the extracted slice as in a brute force search, but only a random selection of them. If the resulting predicate is always the same, the predicate is probabilistically considered a potential OP. This attack can work well for simple numerical OPs with simple slices, but for ours, it will not work on local slices, as it is as infeasible as brute force searching, for the same simple reason that the attacker doesn't know how to limit the space of global memory states to consider as inputs. 

\subsection{Execution Trace Analysis - Fuzzing}
These attacks are specific forms of probabilistic checking, in which the slice is the whole program, and in which the inputs are not selected randomly, but by means of a fuzzer or by letting an attacker select reasonable program inputs. Note that in this attack as discussed by Zobernig et al., only the observed branch directions are taken into account, no further analysis is performed. So the heuristic is that if a conditional branch is only observed to be executed in one direction, we assume it can only be executed in one direction, and its predicate is hence a constant predicate. Flexible one-way OPs can be detected with these attacks in exactly the same way as other one-way OPs. Simply relying on one or a few execution traces will not suffice, however, as that would result in a very large number of false positives and hence in a low precision of the attack: as already observed by Dalla Preda et al~\cite{dalla2006opaque} on CPU benchmarks, in any vanilla benchmark (i.e. without any OPs injected) a large number of conditional branches only execute in one direction when the binary is executed on one or a few representative inputs.

To obtain a clearer, up-to-date view on this aspect, we performed a number of measurements on the real-world test programs we will discuss in more detail in Section~\ref{sec:exp_eval}, and on some additional complex Linux programs (the evince PDF reader and the VLC media player) to confirm this. For example, we traced the Python interpreter on 8 different programs. When an attacker would only consider a single of those 8 traces, between 20\% and 45\% of all possible branch directions would be missed, i.e, of all branch directions occurring at least once in all 8 traces combined. With any combination of 2 out of 8 traces, between 11\% and 36\% of the total would still be missed. To get a decent precision, fuzzing is hence needed. Even then the precision would still be limited~\cite{gan2018collafl,li2017steelix,liang2018fuzzing}, as real-life programs contain many conditional branches that will only be taken in one direction under ``normal'' circumstances, such as NULL-pointer checks following mallocs, or error checks on functions that can actually not really fail, and input validation checks that are either redundant (because they check already checked properties) or not triggered on valid inputs.

Moreover, when the defender does actually worry about this type of attack, the flexible two-way OPs come to the rescue. Be it static ones or dynamic flexible OPs, they offer the same resilience as their fixed counterparts do against this type of attack. To attack those successfully with the described form of execution trace analysis or fuzzing in which only conditional branch outcomes are considered, some form of intra-procedural context-sensitive version of the attack would be needed, albeit with contexts that do not correspond to functions and their callers. While inter-procedural context-sensitive analyses have been proposed to handle obfuscated calls and returns by considering stack frames (a semantical feature not easily obfuscated, unlike the easily obfuscatable call and return instructions)~\cite{10.1007/s10990-011-9080-1}, intra-procedural context-sensitive analysis have to the best of our knowledge not yet been studied. So it remains an open question to what extent they would be useful to detect potential two-way OPs. On our test programs, we did measure that of all conditional branches that were observed to go in two directions in the original, unprotected programs, between 9\% and 18\% is such that the outgoing path only depends on the incoming path. So between 9\% and 18\% of the conditional branches shows the same behavior of static two-way predicates such as the one in the example listings~\ref{lst:protected} and~\ref{lst:protectedtwoway}. This is a strong indication that even context-sensitive execution trace analyses will have limited precision. Investigating this type of complex attacks in more detail is future work. 

\subsection{Pattern Matching}
Dictionary-based pattern matching that can be applied in a plethora of binary analysis use cases: to detect library functions~\cite{flirt}, function prologues and epilogues~\cite{qiao2017function}, high-level programming patterns~\cite{sartipi2003software}, and to deobfuscate mixed-boolean-arithmetic obfuscations~\cite{eyrolles2016defeating}. Zobernig et al.\ observe that surprisingly few different OPs are supported by commonly used obfuscation tools. The dictionaries of patterns to match can hence easily be assembled manually or (semi-)automatically with the help of machine learning techniques~\cite{bao2014byteweight}. By design, flexible OPs that build on APIs already available in the original program, can thwart such attacks to some extent. First and foremost, when a specific API or data structure is used the first time to obfuscate some high-value code, attackers lack the necessary a priori knowledge or samples to construct a dictionary. Secondly, as the re-used APIs are already present in the original program, straightforward pattern matching will offer low precision when defenders deploy them with sufficient stealth, which they can do by carefully designing the encoding of the OPs in the data structures and by deploying them stealthily as described in Section~\ref{sec:stealthy_injection}. There are two caveats, however. First, as pattern matching can be done on a wide range of artifacts at different levels of abstraction, the injection should leave few fingerprints. The  binary rewriter we used for our proof-of-concept implementation lacks in this regard: injected code fragments do not mix naturally enough with the surrounding code. We conjecture, however, that a source-to-source rewriting implementation will do much better in this regard. Studying this in future research would be useful. Secondly, once attackers detect one occurrence of a flexible OP in a (large program) by means of more complex attacks and/or manual effort, they can start searching for the same pattern to find further occurrences. We conjecture that the success and ease of such attack approaches can be severely limited by deploying code diversification techniques on top of the flexible OPs to limit the attack's recall, and by limiting their precision by making sure that the way the predicates are encoded and evaluated resembles operations already present in the original program.

\subsection{Feature Extraction and Machine Learning}
The concept of pattern matching can be generalized to feature extraction. All kinds of features can be extracted from either traces or static code representations, and all kinds of machine learning algorithms can be used both to detect OPs and to deobfuscate them. For detection, the considered labels are ``normal predicate'' and ``opaque predicate''. For deobfuscation, the labels include the types of opaque predicates, and possible the direction in which they are taken, such as ``always true one-way OP''. As noted by Toghifi et al.\ with respect to their supervised-learning approach that considers term frequency features, machine learning approaches suffer from the problem that they need to be trained. They specifically observe that models trained on samples produced with one obfuscator perform badly on code obfuscated by another one~\cite{tofighi2019defeating}. Flexible OPs are therefore as resilient to feature extraction and machine learning attacks as they are to pattern matching attacks. 

\subsection{Taint Analysis}
Zobernig et al.\ use the term taint analysis in a somewhat peculiar way, namely to denote attacks that identify bogus control flow paths corresponding to OPs by detecting a mismatch between the data computations prior to a conditional branch and those following the bogus branch direction. This mismatch can be observed in the form of a lack of (somewhat normal looking) data dependencies, for example when junk code has been injected at the bogus jump target, and that junk code has not been customized to consume the data produced prior to the injected OP. Depending on the objective with which OPs are injected, OPs can be inherently susceptible to this form of attack. For example, in the work of Van den Broeck et al.\ bogus control flow paths are inserted precisely to make disassemblers infer connections between code fragments that are actually not at all related, and thus to thwart the reconstruction of functions~\cite{van2020obfuscated}. As the connected fragments are not related by construction, a certain lack of normal data dependencies is unavoidable. If OPs are inserted simply to complicate the apparent control flow within functions, however, we think it should not be too difficult to choose the insertion locations and the bogus paths in such a way that at there are at least some normal looking data dependencies present. We have not studied this in detail, however, and further research is needed. Compared to fixed obfuscators, our flexible one does provide at least one extra opportunity, however. Prior to an inserted OP, the global data structures are manipulated to set or reset the encoded predicate. Following the OP, be it on a bogus path or a realizable path, the obfuscator can easily inject additional manipulations. In all cases, it can inject the orthogonal manipulations as described in Section~\ref{sec:inject_support}, and in cases where the predicate is not live after it has been checked for the OP, even set and reset manipulations can be injected. Code prior to and following the OP and its branch will therefore at least operate on the same data, this potentially making this form of attack harder. 

\subsection{Automated Proving}
One can try to let a SMT-solver prove that a computed predicate is a constant. As noted by Zobernig et al., automated proving on concrete code slices is only feasible for sufficiently simple OPs, with sufficiently small input domains. The flexible OPs on top of complex data structures and global state that we proposed are not in the scope of such automated proving. 

\subsection{Symbolic Execution}
SMT-solvers also back symbolic execution engines~\cite{king1976symbolic}. Ming et al.\ showed that simple OPs can be detected by symbolic execution~\cite{ming2015loop}. Recently, Banescu et al.\ evaluated some state-of-the-art symbolic execution engines on, amongst others, OPs~\cite{banescu2016code}. They observed that OPs have a small impact on the slowdown of symbolic execution. They also observed that when the injected OPs do not insert input-dependent computations, the opaque branches are easily eliminated. Yadegari et al.\ also demonstrated that deobfuscation by means of symbolic execution is possible~\cite{yadegari2015symbolic}. To counter such attacks, transformations have been proposed that lead to path explosion, such as range dividers~\cite{banescu2016code} and split/for/write-obfuscations~\cite{obf_anti_symbolicdeobf}.

To attack OPs or other obfuscations in binaries, such as mixed-boolean-arithmetic (MBA) data obfuscation~\cite{zhou2007information}, symbolic execution is typically deployed locally, i.e., on a small program fragment or slice~\cite{eyrolles2016defeating}, for example a path within a function leading up to a suspected OP conditional branch. Well-known binary symbolic execution tools such as \texttt{BINSEC}~\cite{djoudi2015binsec,binsec} and \texttt{angr}~\cite{angr} offer this functionality. Such analyses have been integrated in popular reverse-engineering tools such as IDA Pro~\cite{idapro}, Binary Ninja~\cite{binaryninja} and Ghidra~\cite{ghidra}, for example in the DROP tool that integrates the attack of Ming et al.\ into IDA Pro~\cite{DROP}, such that an attacker can interactively invoke the attack at manually chosen program points or code fragments. In the case of OPs, the attacker aims to let the symbolic execution engine's SMT-solver solve path conditions on the selected fragments that lead to the execution of different outgoing paths of the OP. Similar to how MBA is attacked~\cite{eyrolles2016defeating}, the solver can be customized to target known OPs from fixed obfuscators by means of template-based normalization rules for path conditions, i.e., rules that recognize and simplify complex path conditions corresponding to known OPs.

Symbolic execution engines can be considered smarter attacks than ``dumb'' brute force search, probabilistic checking, and execution trace analysis. Rather than (randomly) trying many potential inputs to produce the relevant behavior, the SMT-solver tries to find inputs in a smart, targeted manner by solving path conditions. Still, the smart attacker and the dumb attacker suffer from exactly the same problem when facing our flexible OPs that build on complex data structures and global state. As discussed above, the attackers and their engines do not know what inputs (including what memory states) can or cannot occur at the start of the symbolically executed slice.

To validate our assessment of how symbolic execution fares with our flexible OPs, we used the popular tools \texttt{angr} and \texttt{BINSEC} to attack function \texttt{A} in Listing~\ref{lst:protected} in the program version without function \texttt{E}, i.e., where the OP on line 10 truly is a one-way OP. Specifically, the tools symbolically executed that function, starting at its entry point on a symbolic input memory state, i.e., every possible memory state. In about 10 seconds, \texttt{Angr} found two potential input states for \texttt{A} that make the OP go in both directions. \texttt{BINSEC} did the same in about 20 minutes. Aside from the difference in required run time, both tools hence yield the same results that the predicate might evaluate to both values, and hence fail to identify it as a one-way OP. The reason is of course that the symbolic execution engine in this case is not limited to memory storing disjunct graphs, which the program is. Operating on binary code that lacks type information, the symbolic execution engine and the SMT-solver consider any combination of pointers that lets the predicate evaluate in either direction, and they easily produce such combinations. The only solution for an attacker then is to let the attacker specify what the possible input states could be. One option is to use concolic execution, and actually let the symbolic execution start from a memory state that was observed in a real execution. This, however, runs into the exact same problem that execution trace analysis had to face, namely the problem of ensuring that sufficient coverage is achieved with the selected inputs.

Even if attackers can overcome that problem, users of the flexible obfuscators have additional options to thwart symbolic execution. First, resilience can be achieved by encoding the predicate values in ways that are not known a priori to attackers, such that they cannot exploit template-based normalization rules for path conditions. This is, obviously, only possible for flexible obfuscators. In addition, the user of a flexible obfuscator has the option to embed concrete path-oriented protections, such as the aforementioned range dividers and split/for/write-obfuscations, in the code fragments that the obfuscator will inject, thus causing path explosion leading to inpractical symbolic execution running times. In general, all the existing defenses against symbolic execution can be included in the code fragments that will be injected and that operate on the complex data structures.

In conclusion, our appoach offers the potential to inject resilient OPs, and it is up to the user of the flexible obfuscator to decide which level of resilience is to be achieved, and to design the used datastructures, the encoded invariants, and the code fragments to be injected accordingly.

\subsection{Abstract Interpretation}
Cousot presented abstract interpretation, a general framework for program analysis~\cite{cousot1977abstract}. Dalla Preda et al.\ used abstract interpretation to detect (relatively simple) one-way OPs~\cite{dalla2006opaque} and they proposed to measure strength of OPs by means of abstract interpretation~\cite{dalla2009semantics}. An underlying assumption of their work makes it practically useless on flexible OPs. Specifically, they assume that the attacker can extract a slice easily and that the attacker knows the abstract domain with which the extraced slices need to be analyzed. With flexible OPs, however, the attacker cannot know the domain a priori, just like he doesn't know what patterns to try to match, or what template-based normalization rules to use. 

\subsection{Static Analysis}
Simple OPs can also be detected and deobfuscated through static data and control flow analyses that come down to abstract interpretations, but are typically not handled as such by attackers or in the literature. For example, using value set analysis~\cite{balakrishnan2004analyzing}, the Binary Ninja disassembler can determine that simple implementations of algebraic OPs such as $(x^2-x)\, mod\,2$ evaluate to constants. With simple implementations, we mean implementations in which $x$ is not read from memory or dependent on an input parameter of a function, because in those cases, Binary Ninja's analysis lacks the necessary precision. It can therefore certainly not handle the flexible OPs we propose. Fundamentally, all static analyses come with limited precision, if only because memory alias analysis is an NP-hard~\cite{aliasNPhard}. To attack our flexible OPs on complex data structures with global state, an interprocedural analysis is required that offers high enough local precision. We conjecture that this is practically infeasible for all but the simplest uses of flexible OPs. One reason is that the problem is aggrevated by the reuse of API functions used in the original program. Because those functions are used in multiple, completely independent calling contexts in a protected program, in many cases it will not be possible to compute simple pre- or post-conditions that hold for all of a function's inputs or outputs. In other words, simple domains will not suffice, and what domain should be used is, as in the case of abstract interpretation, not known a priori.

Moreover, to achieve sufficient precision, designers of analyses typically combine different forms of sensitivities (to paths, flow, fields, allocation-sites, contexts, etc.), which they deploy at certain depths, such as K-depth context-sensitivity in which each function's considered calling contexts are call chains of length K. For some analyses, they also choose so-called summary equations to model the data flow through functions, for example modeling the return value of a function as a first-order linear combination of its inputs. If for some more complex function the chosen form of summary equation is not expressive enough to capture the semantics of the function conservatively, the summary equation defaults to ``don't know'', resulting in a loss of analysis precision. In general, relying on NP-hard problems such as aliasing to evaluate the resilience of software protections is not correct, because even if some problem is NP-hard in theory, it is typically very difficult to generate instances of the problem that can be injected as a software protection and that cannot be handled by practical and precise enough analysis algorithms. Indeed, when the problem to tackle is known, algorithm designers can often come up with the exact combination of sensitivities, depths, and summary functions to achieve the required precision. For example, Udupa et al.\ clone basic blocks in flattened control flow graphs to create sufficient analysis contexts for them~\cite{Udupa}. To select the blocks to be cloned, they specifically build on the a priori knowledge of flattening dispatchers having many incoming and outgoing edges. For cases where the necessary a priori knowledge is lacking to select blocks to clone, they offer no concrete solution.

In the case of flexible OPs, attackers don't know upfront whether their analysis needs to tackle trees, sets, hashtables, or any other form of complex data structure. Neither do they know which invariants or other features they can aim for. For these reasons, we conjecture that flexible OPs, when building on sufficiently complex data structures, will be very resilient against static analyses. 

\subsection{Mutation Attacks}
In theory, attackers can mutate instructions in the program to check if those mutations effect the program as it executes on chosen inputs. Such attacks may be used to detect dummy code or junk code inserted on bogus control flow paths. Zobernig et al.\ note that such attacks are hard to automate and scale to larger programs, however. They hence are of little practical value. Moreover, unlike in the simplified example in Listing~\ref{lst:protected}, the uses of one-way and two-way opaque predicates we put forward do not involve junk code at the tail of bogus control flow transfers. So mutation attacks are moot for these uses. 

\subsection{Trace Simplification}
Several deobfuscation techniques have been proposed in recent years. Generic deobfuscation by Yadegari et al.~\cite{yadegari15generic} identifies and simplifies quasi-invariant operations in traces as well as operations that according to a bit-level taint analysis~\cite{bitlevel_taint} do not contribute to the observed IO-relation. Two-way OPs are resilient to the simplification of quasi-invariant behavior because they do not behave as such by definition. Furthermore, if the existing API functions that are reused in the flexible protection were not only present but also executed in the original program, which the defender can easily check with a profiler or a tracing tool himself, also their code will not behave quasi-invariantly, as it is executed in multiple contexts in the protected program. In addition, by using data that is relevant and input-dependent in the original program as inputs to the injected fragments, in particular the initialization code of the global data structures, the used taint analysis will fail to mark the operations on the complex global data structures as non-contributing to the programs IO-behavior. In Listing~\ref{lst:protected}, the initialization of \texttt{X} as a deep copy of \texttt{G}, and the dependency of \texttt{temp1} on an input file examplify this. This feature also makes the flexible OPs resilient to other taint-analysis based attacks~\cite{wang2008still,seead}. Moreover, a user of the flexible obfuscator can deploy known anti-taint techniques, such as side channels~\cite{sarwar2013effectiveness} and implicit information flow~\cite{cavallaro2008limits,stephens2018probabilistic}, in the injected code fragments, to provide further hurdles to taint-analysis-based techniques such as generic deobfuscation. 

Synthetic code generation is another recently proposed deobfuscation technique~\cite{blazytko2017syntia}, in which the semantics of a complex, obfuscated instruction sequence extracted from a complete program trace is replaced by a simpler, synthesized code fragment that produces as much as possible the same outputs as the original fragment for a large set of randomly selected inputs. The authors of the technique deploy it on clearly delimited instruction sequences, such as those corresponding to one MBA-expression, one instruction handler in a bytecode interpreter, or one Return-Oriented-Programming (ROP) chain. They first execute those fragments on the set of randomly selected inputs to determine the expected, corresponding outputs. Their code synthesis process is then guided by Monte Carlo Tree Search (MCTS). It iteratively grows the complexity of the syntesized code in the direction that produces the expected outputs for as many as possible of the selected inputs. Our proposed OPs are resilient against this form of deobfuscation for several reasons. Most importantly, like local symbolic execution, brute force search, probabilistic checking, and execution trace analysis, the attacker lacks the knowledge to select only those random inputs that can actually occur during an actual program execution. When the attacker also selects inputs that correspond to unrealizable global states that would make the OP evalute in the bogus direction, the code synthesis will aim for code that also implements the semantics of the bogus path, and then hence still includes a predicate and a conditional branch. Moreover, it is unclear how an attacker would be able to select the instruction fragments to be extracted and deobfuscated, given that the slices contain calls to functions also used elsewhere in the code, and since the injected fragments operate on global state, with a long lifetime. In short, we don't consider this technique viable to attack our flexible OPs. 

\subsection{Devirtualization}
A specific case of trace simplification concerns devirtualization. Coogan et al.~\cite{deobf_virtobf} use similar techniques as Yadegari et al.~\cite{yadegari15generic}, against which our flexible OPs are resilient as discussed above. Other devertualization techniques aim for abstracting a virtual program counter, which they do by exploiting features typical for virtualization-based obfuscation~\cite{Kinder2012,reverseng_malwareemu} and the resulting traces. Adapting those abstraction techniques to target flexible obfuscations is comparable to trying to use abstract interpretation to attack flexible OPs without a priori knowledge about the abstraction domains to be used. We hence conjecture that engineering abstraction techniques similar to those used in devirtualization, and then reusing them to deobfuscate multiple programs protected with different OPs is not feasible with the current state of the art.

\section{Experimental Validation}
\label{sec:exp_eval}
In the previous section, we performed an extensive theoretic security analysis, backed up with a number of measurements on real use cases and experiments on small samples to provide ample evidence of the resilience and stealth of the proposed use of flexible OPs.

In this section, we evaluate their cost and feasibility by means of a proof-of-concept implementation and real-world use cases. 

\subsection{Proof-of-concept Implementation and Experimental Setup}
We implemented support for one-way and two-way flexible OPs in Diablo, a post-link-time binary rewriting research tool~\cite{diablo2005} that already includes support for a range of other MATE protections including control flow obfuscations~\cite{van2020obfuscated} and diversification~\cite{coppens2013feedback}. We opted for Diablo because we can reuse the basic functionality for those protections, hence limiting the engineering effort for developing this proof-of-concept implementation. We do not consider a link-time rewriter such as Diablo, which does not lift the code to a more abstract IR but instead transforms assembly-level code, a good choice in terms of security for real-world deployment, however. Rewriting code at such a low level of abstraction simply leaves too many fingerprints, as already mentioned in the previous section. For real-world deployment, a source-to-source rewriting obfuscator would likely be a much safer choice. As this experimental evaluation focuses on the practical feasibility of the proposed flexible OPs rather than what is the best way to inject them, we think Diablo suffices for this evaluation. The code for our implementation is available as open source at \url{https://github.com/csl-ugent/diablo/tree/meta-api}.
 
The targeted platform of our proof-of-concept is Linux ARMv7. We performed measurements on a SABRE Lite i.MX development board running Ubuntu 18.04.3 LTS (kernel 4.9.88-boundary-14b)~\cite{imx6}. The use cases were built with GNU GCC 5.5.0 (hard-float) and binutils 2.26.1.

\subsection{Benchmark Samples}
We evaluated our proof-of-concept implementation on four open-source use cases. The support code, including the four meta-API descriptions, is available online at \url{https://github.com/csl-ugent/meta-api}. All flexible OPs defined below can be used both as contextual OPs and as flow-insensitive invariant OPs. 

\textbf{Python} Our first use case is the interactive shell (i.e., the interpreter) of the reference implementation of the Python programming language (v3.7.4)~\cite{python}. This program is written in C. Invoked on Python programs, it compiles them into intermediate bytecode and executes them in a virtual machine. As inputs for our measurement experiments, we invoked Python on six usage scenarios: installing its package manager PIP, using that to install 3 packages, and running those packages on sample inputs provided by their respective developers.
For the flexible OPs, we reuse the \texttt{\_Py\_Hashtable\_t} data structure and its APIs. We encode OPs in two predicates: P1 is true iff a given key is present in a global hashtable, P2 is true iff two given keys are present with the same value. Because P2 requires double the amount of operations on the hashtable to set/reset/check the predicate, its use is about twice as expensive in terms of execution time overhead. To provide stealthy keys, the injected code reuses strings already present in the original program.

\textbf{KeepassXC}
KeepassXC (v2.6) is an open-source password manager that stores different types of information in a tree-like structure~\cite{keepassxc}. It builds on the Qt framework and is written in C++. For our measurements, we performed numerous operations (spread over 21 different runs) to create and manipulate a database, look up data and manage users. We then first used the meta-API to encode two flexible OPs reusing Qt's \texttt{QString} class: P1 is true iff two given strings are equal, P1' is true iff a give string ends with a certain suffix. P1 and P1' have similar complexity and performance overhead. Secondly, we encoded a flexible OP in a global \texttt{Group} data structure that represents a directory tree for storing passwords. P2 is true iff the directory tree representation contains a given file in a given path. To increase the stealth of the injected API calls, we reuse four strings that are already present in the original program. The use of P2 is two orders of magnitude more expensive in terms of execution than that of P1 and P1'.

\textbf{Ninja}
Ninja (v1.10.0) is an open-source software build tool that provides a faster alternative for the well-known automake tool~\cite{ninja}. The perform measurements, we let Ninja, which is written in C++, build itself, letting the main command-line utility perform a dry-run build. Ninja already makes use of the C++ STL sets and maps, which are both template-specialized versions of the STL's private red-black tree implementation class \texttt{\_Rb\_tree\_impl}. We reused the latter tree class to encode three flexible OPs on global trees. P1 is true iff the tree contains a given value. P2 is true iff two given nodes are part of the same tree. P3 is true iff all nodes in the given tree have an even value. P1 is cheapest for checking the value of the predicate, while our implementation for P2 is an order of magnitude cheaper for setting/resetting the predicate. P3 is in between in our implementation. P1 is hence less costly for flow-insensitive invariant OPs, while P2 is less costly for contextual OPs.
To increase the stealth of these flexible OPs, we instructed the obfuscator to generate small numbers to feed to the called APIs. Those small numbers can be encoded in single constant-producing instructions, which also occur frequently in the original program.

\textbf{Radare2}
The Radare2 framework (v3.8.0) provides a set of command-line utilities for reverse engineering binaries~\cite{radare2what}. For this use case, we did not protect a main binary, but the dynamically linked shared \texttt{util} library. To perform measurements, we performed 12 common binary analysis operations on Linux' \textit{ls} utility, for which the interactive shell of the main Radare2 utility was invoked. The operations include looking up the program's entry point, auto-analysing the binary, listing the imported functions and program strings, computing the cross-references for some string, listing the functions in the binary, disassembling the main function and rendering an SVG file for that function.
We used the meta-API to prescribe three flexible OPs on Radare2's internal graph implementation in class \texttt{RGraph}. This class also served as the inspiration for the example in Listing~\ref{lst:protected}. P1 is true iff a certain path is present in a given graph, P2 is true iff two given nodes are part of the same graph, and P3 is true iff the give graph contains a one-edge loop. Our implementation of P1 is slightly cheaper when it comes to checking the predicate, whereas P2 is a factor seven cheaper for setting/resetting the predicate.

For each use case and each encoded predicates, we used the meta-API to prescribe one evaluation code fragment. For each use case, we also selected one predicate for which we provided more than one setter and more than one resetter fragment, thus giving the flexible obfuscator more freedom to select set/reset and evaluation combinations to inject contextual OPs. 

With these use cases and the range of data structures we reuse in them, we demonstrate the large variability of flexible OPs that can be deployed in practice. How to encode OPs is really only limited by the imagination of the flexible obfuscator's user, and of course by the available data structures to reuse and the cost of the injected OPs. The latter is evaluated in Section~\ref{sec:eval_cost}.

To validate the correctness of our proof-of-concept implementation, we sprankled the four use cases with one-way and two-way OPs, randomly injecting OPs in up to 50\% of the executed basic blocks in their code and checking the correct execution of the thus transformed programs (on the above described inputs) until we no longer triggered deviations in the program's input/output. This level of validation suffices to trust the results of the cost evaluation below. 

\subsection{Cost Feasibility Evaluation}
\label{sec:eval_cost}
As already mentioned in the previous section, some injected operations to set/reset/check the encoded predicates are much more expensive than others. Moreover, even the cheaper ones that we implemented are pretty heavy-weight compared to OPs previously proposed in the literature. Sprinkling a program with expensive operations such as directory look-ups, tree searches, string comparisons, etc.\ to obfuscate most of those programs' control flow is not likely feasible.

Stealth is another reason for deploying the heavy-weight flexible OPs sparsely. If a whole program is sprinkled with the same or similar constructs and API calls, their frequent occurrence alone might make them suspicious to attackers. The defender needs to co-design the flexible OPs with the application to be protected, which makes using flexible OPs much more costly in terms of manual effort than fixed OPs, and the number of data structures to be reused is inherently finite for any program. The defender's investment has to be protected, in part by deploying the designed flexible OPs in such a way that their cover is not easily blown. In other words, it makes sense to deploy flexible OPs only in those parts of a program where strong protection has the highest priority.

This observation then begs the question as to where in a program sparse deployment of some specific flexible OPs is feasible, given the defender's OP design and his performance overhead budget. To answer this question, a defender can use the following strategy, which we also applied to the four use cases. 

First, we have measured the average execution times of the implemented flexible OP operations, i.e., of set, reset, and check operations. We did so by executing various combinations of them in microbenchmark loops injected into the use cases close to their entry points. This process can be automated completely in the flexible obfuscation tool, and hence requires no additional manual effort of the defender.

Secondly, we profiled the use cases to collect basic block execution counts and we measured their execution time, using the inputs discussed in the previous section.

From the collected information, one can easily compute an estimate of how expensive it would relatively be to inject an OP at some program point. In any case, it requires the execution of a number of predicate evaluation operations equal to the execution count of the program point. In the case of contextual predicates, a basic implementation like the one in our proof-of-concept tool additionally requires that one set or reset operation is executed per executed evaluation operation. For both flow-insensitive invariant and contextual predicates, it is hence easy to estimate how much additional execution time would be incurred by the injection of a selected OP at any particular program point, and hence what the relative slowdown would be. The same can be done for the injection of additional set, reset, or orthogonal data structure operations. 

We validated this approach on the use cases by randomly deploying OPs in the hottest parts of the use cases and measuring their overhead, and we confirm that the estimates where within a couple of percentages from the measured overheads. 

\begin{figure}[t!]
  \centering
  \includegraphics[]{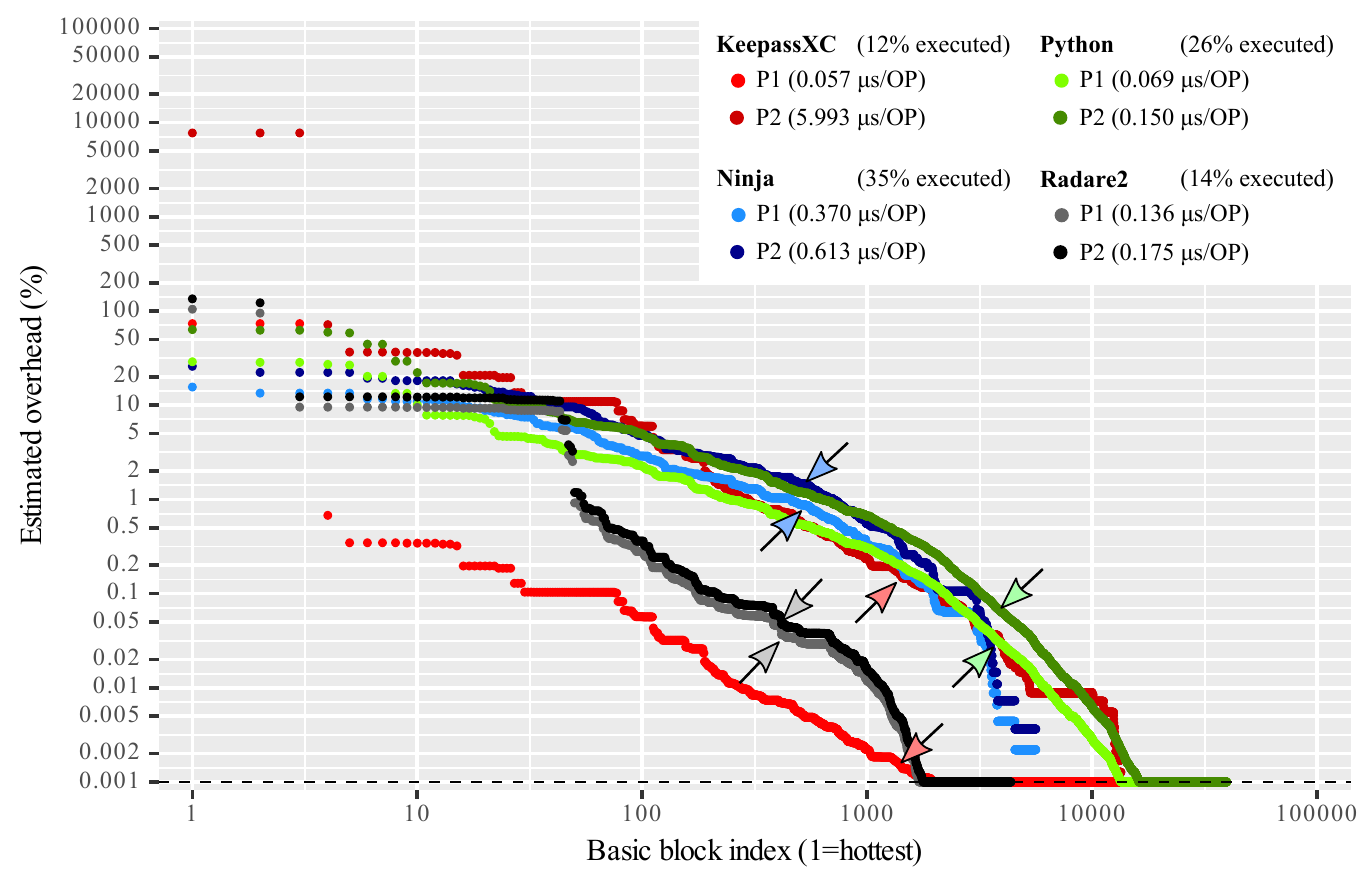}
  \vspace{-0.9cm}
  \caption{Estimated overhead for flow-insensitive invariant OPs requiring only an evaluation per executed OP.}
  \label{fig:invariant_cost}
  \Description{Fully described in the text.}
  \vspace{0.7cm}

  \includegraphics[]{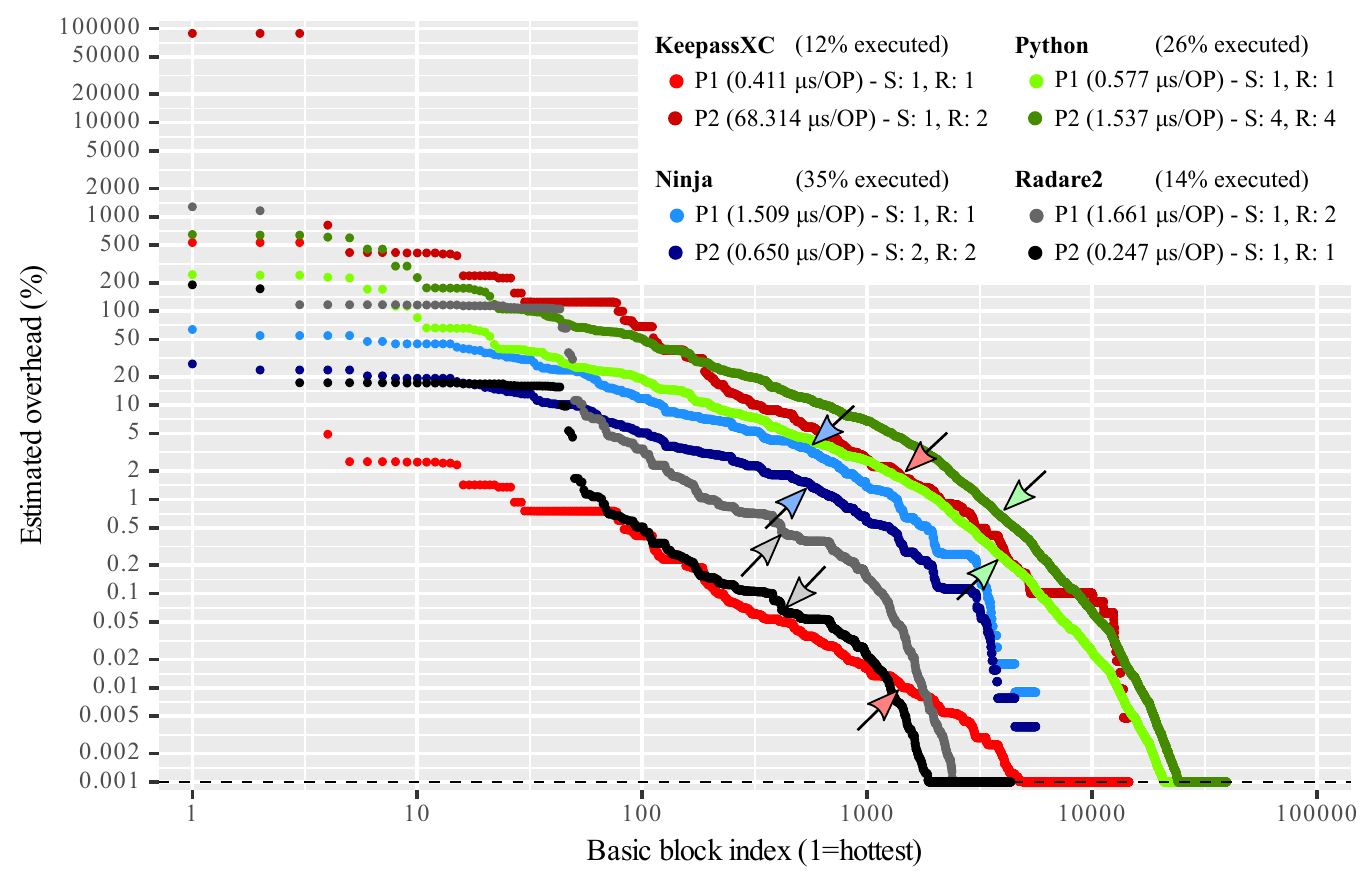}
  \vspace{-0.9cm}
  \caption{Estimated overhead for contextual OPs requiring a set/reset and an evaluation per executed OP.}
  \label{fig:contextual_cost}
  \Description{Fully described in the text.}
\end{figure}

Figures~\ref{fig:invariant_cost} and~\ref{fig:contextual_cost} plot the thus estimated relative overhead per basic block in each use case, for the cheapest and the most expensive OPs (i.e., P1 and P2 as introduced above for the use cases). The basic blocks are ordered from most frequently executed to least frequently executed. Blocks that were not executed for any of the discussed inputs are excluded. We exclude them for the sake of clarity of the graphs and also because we deem defenders to be likely not interested in obfuscating code that is never executed (at least, not executed for the considered inputs). The percentage of all blocks that are included is indicated in the legends of the charts, as is the average execution time per executed OP. These execution times are averages for the different set/reset/evaluation code fragments we provided with the meta-API per predicate. The R/S/E numbers in the legends indicate exactly how many of them we provided. Overheads lower than 0.001\% are rounded off to 0.001\%, hence the straight lines at the left bottom of both charts.

Some interesting observations can be made from these charts. First, the data confirm that invariant OPs are much cheaper in most cases, up to an order of magnitude in some cases. Secondly, it is clear that each program has a number of hot program points that are executed simply too frequently to be feasible injection points for flexible OPs. Even for invariant OPs injecting the cheapest OPs into the hottest points in a program would yield run time overheads between 16 and 7765\%.

What colder fraction of all executed blocks can be protected with acceptable overhead can be determined from the individual lines on the charts. As an example, the arrows in the charts mark the first basic block in each benchmark that is not within the benchmark's 10\% hottest blocks. In other words, for 90\% of the executed blocks of a program, the overhead of injecting one OP in them will be lower than the Y-value marked by each arrow. For some benchmarks and particular OPs, the overhead of injecting a flexible invariant OP in those 90\% is below 0.1\%, meaning that with a 10\% overhead budget, at least 100 OPs can be injected in the 90\% coldest executed code. That implies that a significant fraction of the colder security-sensitive functionality can indeed be protected with flexible OPs. For other benchmarks, the overhead drops below 0.1\% only for a smaller percentage of the coldest executed code.

Overall, these results confirm our earlier argument that flexible OPs should only be used sparsely to protect the most security-sensitive code parts. Moreover, it is clear that at least flexible OPs with the complexity of our designs for the use cases are not applicable to the hottest code parts because of the large overhead they incur.

Charts like the ones in Figures~\ref{fig:invariant_cost} and~\ref{fig:contextual_cost} can easily be generated automatically for a flexible OP design and use case. The defender only needs to provide the necessary inputs. Furthermore, if the defender marks the security-sensitive code fragments, the charts can be filtered for the basic blocks in those fragments. With those charts, the defender can then easily evaluate the feasibility of his flexible OP designs. While the use of such charts as a decision support tool might not suffice to find designs of sufficient complexity to provide protection and of low enough complexity to be applicable to the hottest security-sensitive code fragments, they will make the life of the flexible obfuscator user much more productive, thus easing the burden of using flexible obfuscation.

\section{Related work}
\label{sec:related_work}
To the best of our knowledge, we are the first to pitch the concept of flexible obfuscation tools, and nothing close to the techniques we presented in this paper for flexible obfuscation exists already. For related work on OPs, we refer to the discussion in Section~\ref{sec:opaque}. For related work regarding attacks on and analysis of obfuscated programs, we refer to Section~\ref{sec:evaluation}.

\section{Conclusions}
\label{sec:conclusions}
In this paper, we proposed the novel concept of flexible software protection tools that do not hardcode fixed implementations of software protections but instead implement them through the reuse of data structures and APIs already available in the software to be protected. This decreases the learnability and hence the a priori knowledge of attackers about the protection implementations, allows using global state encoded in complex data structures, and avoids quasi-invariant and semantically irrelevant behavior of relevant code fragments, thus increasing the resilience and stealth of the protections under a wide range of attacks.

We concretized the concept for opaque predicates, one of the most essential building blocks of software protection against man-at-the-end attacks and reverse engineering. We presented a meta-API that builds on fundamental concepts such as pre- and postconditions with which users can prescribe how existing data structures and their APIs can be reused by the flexible obfuscator. We performed an extensive theoretical security analysis, backed up with concrete measurements on real-world applications and experiments in which a symbolic execution attack failed on even small toy examples. We developed a proof-of-concept flexible obfuscator, and deployed it on four real-world programs to study the feasibility and cost of using flexible opaque predicates.

The overall conclusions of these experiments and discussions are that flexible OPs can provide strong resilience and improved stealth, but also that they are costly and hence to be used only sparsely, on code fragments that do not dominate performance.

\begin{acks}
This research was funded by the \grantsponsor{IWT}{Agency for Innovation by Science and Technology in Flanders (IWT)}{} under Grant No.~\grantnum{IWT}{141758}. This research was also partly funded by the Cybersecurity Initiative Flanders (CIF) from the Flemish Government and by the \grantsponsor{FWO}{Fund for Scientific Research - Flanders (FWO)}{} under Project No.~\grantnum{FWO}{3G0E2318}.

The authors would like to thank S\'ebastien Bardin and Sebastian Banescu for sharing their insights on the resilience against symbolic execution, Fr\'ed\'eric Recoules for evaluating the example program on BINSEC, and Thomas Van Cleemput for for performing embryonic experiments on our very first ideas.
\end{acks}

\bibliographystyle{ACM-Reference-Format}
\bibliography{journal}

\end{document}